%% file: WNM.tex
\documentclass[onecolumn,showpacs,aps,superscriptaddress,showpacs,prc,eqsecnum]{revtex4}
\usepackage[dvips]{graphicx}
\usepackage[english]{babel}
\usepackage{here}
\usepackage{subfloat}
\usepackage{fontenc}
\usepackage[utf8]{inputenc}
\usepackage[colorlinks=false]{hyperref}
\usepackage[printonlyused]{acronym}
\usepackage{bm}
\usepackage{bbold}
\usepackage{amsmath}
\usepackage{slashed}
\usepackage{amssymb}
\usepackage{textcomp}
\usepackage[usenames]{color}
\usepackage{enumerate}
\usepackage{dcolumn}
\usepackage{longtable}
\usepackage[usenames]{color}
\include{definitions}

\begin{document}

\title{Thermal Properties of Asymmetric Nuclear Matter}
\author{A.~Fedoseew} \affiliation{ Institut f\"ur Theoretische Physik,
Universit\"at Gie\ss en Heinrich-Buff-Ring 16, D-35392 Gie\ss en, Germany }
\author{H.~Lenske} \affiliation{ Institut f\"ur Theoretische Physik,
	Universit\"at Gie\ss en Heinrich-Buff-Ring 16, D-35392 Gie\ss en, Germany }
	\affiliation{ GSI Helmholtzzentrum f\"ur Schwerionenforschung, Planckstra\ss
	e 1, 64291 Darmstadt, Germany }
\date{\today} 
\begin{abstract} The thermal properties of asymmetric nuclear matter are
	investigated in a relativistic mean-field approach. We start from free space
	NN-interactions and derive in-medium self-energies by Dirac-Brueckner theory.
	By the DDRH procedure we derive in a self-consistent approach
	density-dependent meson-baryon vertices.  At the mean-field level, we include
	isoscalar and isovector scalar and vector interactions. The nuclear equation
	of state is investigated for a large range of total baryon densities up to the
	neutron star regime, the full range of asymmetries $\xi=Z/A$ from symmetric
	nuclear matter to pure neutron matter, and temperatures up to $T \sim
	100$~MeV. The isovector-scalar self-energies are found to modify strongly the
	thermal properties of asymmetric nuclear matter. A striking result is the
	change of phase transitions when isovector-scalar self-energies are included.
\end{abstract} \maketitle
%
\section{Introduction}\label{sec:Intro}

A major task of nuclear many-body theory is to understand the equilibrium
properties of nuclear matter under variations of density, pressure,
proton-to-neutron fraction, and, last but not least, temperature. For this
demanding goal decade-long experimental and theoretical efforts have been made,
ranging from early studies of compound nuclei and highly excited pre-compound
systems to studies of exotic nuclei at extreme isospin and compressed baryonic
matter in high energy heavy ion collisions. Considering the status of research
on the nuclear equation of state (EOS) one finds that many theoretical studies
exist for symmetric and pure neutron matter. The work of Sauer \emph{et al.}
\cite{Sauer:76} is probably the first systematic study to derive the thermal
properties of finite nuclei on theoretical grounds. Actually, historically
nuclear matter studies were strongly motivated by astrophysical issues
\cite{Fried:81}, with a especial need for investigations of asymmetric matter
with arbitrary values of the charge asymmetry defined in terms of the proton
fraction $\xi=\frac{Z}{A}$, ranging from $\xi=\frac{1}{2}$ in symmetric matter
to $\xi=0$ for pure neutron matter. Thermodynamically, symmetric nuclear matter
and pure neutron matter, respectively, correspond to to single fluid systems.
Since in symmetric nuclear matter isovector self-energies are absent by symmetry
reasons, protons and neutrons are dynamically indistinguishable, hence forming a
single-component fluid. The one-component character of neutron matter is
obvious. The theoretical description of such one-component quantum systems is
much simpler than that of a multi-component fluid. As is well known and will be
seen also in later sections of this work, the treatment of asymmetric nuclear
matter requires first of all extended theoretical methods and secondly such a
two-fluid systems shows new features which are absent (or hidden by symmetry
reasons) in a single-fluid nuclear matter.

Equilibrium thermodynamics, as pursuit here, is the appropriate approach to the
nuclear equation of state as a function of density and temperature. Known
essential properties are the liquid-gas phase transition at sub-saturation
densities and moderate temperatures. An important question is how the system
evolves by passing various \emph{binodals}, denoting phase separation
boundaries, and \emph{spinodals}, indicating stability or, likewise,
instability boundaries. Studies of infinite and finite nuclear systems indicate
that symmetric nuclear matter undergoes a liquid-gas phase transition at
critical temperatures in the range of $T_C=10\cdots20$~MeV
\cite{Fried:81,Baldo:99}. The results depend, however, on the chosen $NN$
interaction. Phenomenological density functional models seem to favor lower
$T_C\sim 10$~MeV, as in the very early work of Sauer \emph{et al.}
\cite{Sauer:76} while values around $T_C\sim 20$~MeV are predicted by
microscopic approaches as, for example, in the study of Baldo and Ferreira
\cite{Baldo:99}. Covariant field theory and thermodynamics have been studied
very frequently. A comprehensive discussion is found in the early work of
Weldon \cite{Wel:82}. The connection of a hadron field theory and
thermodynamics was discussed by Furnstahl and Serot \cite{Furn:90}. Finite
temperature many-body theory, the Green functions, and the solution of the
G-matrix equation in a transport theoretical connection has been reviewed in
detail by Botermans and Malfliet \cite{Bot:90}. The same authors have studied
intensively in-medium interactions in cold and hot nuclear matter
\cite{terH:86,terH:87}. As reported in \cite{terH:87} they found that the
Dirac-Brueckner G-matrix depends only very weakly on temperature. As will be
discussed later, this allows to extrapolate interactions safely from the $T=0$
to the $T>0$ case. The work of M\"uller and Serot \cite{Muel:1995} addresses
the thermal properties of asymmetric nuclear matter quite generally. On the
basis of a relativistic mean-field model with non-linear self-interactions of
the scalar and vector fields and relativistic thermodynamics, the phase
structure of nuclear matter with arbitrary proton content was investigated in
detail. An unexpected result was that in asymmetric matter instabilities are
induced primarily by fluctuations in the proton content rather than by
fluctuations in the net baryon density. Hence, \emph{chemical instability} of
the composition of the fluid wins over the \emph{mechanical instability} of the
total system, at least in that particular model. The instability region of warm
asymmetric nuclear matter has been studied by many other groups.  A detailed
comparison between relativistic models and Skyrme forces is provided by the
work of Dutra et al \cite{Dutra:2008}.

The mentioned properties are largely determined by the $NN$-interactions,
besides, naturally, Fermi-Dirac statistics. At the energy and momentum scales,
relevant for nuclear matter, $NN$-forces are acting like van-der-Waals forces
among molecules. Hence, at not too high density, pressure, and temperature
nuclear matter resembles a multi-component van-der-Waals gas. However, the exact
values of the permissible density and temperature ranges are yet to be
determined. It is one purpose of this work to add new aspects to that ongoing
research program. As far as temperature is concerned, the QCD-phase transition
into a Quark-Gluon-Plasma ($QGP$) at $T^{QCD}_c\sim 150$~MeV is a clear limit.
Hence, only temperatures well below $T^{QCD}_c$ should be considered. We choose
a temperature range $T<100$~MeV. The density behavior is much less constraint.
The assumed central density  of a neutron star may serve as guideline. Taking
into account the recently observed neutron star with a mass of twice the solar
mass, we may accept a density range below $6\rho_{eq}$, where
$\rho_{eq}=0.16fm^{-3}$ is the density of symmetric nuclear matter density at
the saturation point.

In addition to the liquid-gas phase transition, equilibrated nuclear matter may
undergo other phase transitions by creating hadrons. Pion and kaon condensation
has been investigated theoretically, and also the thermal production of excited
baryons like the $\Delta(1232)$ resonance were considered. However, clear
experimental signals indicating such hadron condensation with the sudden
appearance of a new type of species in the fluid are still missing. Better
established are bosonization processes in cold matter at extremely low
densities: under such conditions nuclear matter rearranges into bound
constituents, forming deuterons and, especially, $\alpha$ particles.  In the
crust of neutron stars, one expects a lattice of heavier nuclei up to the iron
and nickel region \cite{ShaTeu:1983}. Finite temperature effects on the
geometrical formation of the neutron star crust have been analyzed in detail by
G\"ogelein et al. \cite{Goegelein:2008}. Here, we consider pure uniform nuclear
matter, composed of nucleonic quasi-particles only but with an arbitrary
mixture of protons and neutrons.  A very complete and comprehensive discussion
of thermal dynamics in asymmetric nuclear matter is found in the already cited
work by M\"uller and Serot \cite{Muel:1995}. These authors have pointed out in
remarkable clarity the important differences between the thermodynamics of
symmetric and pure neutron matter on the one side and the thermodynamics of
asymmetric matter. Symmetric and neutron matter as single component systems
with a single conserved charge, namely baryon number, are corresponding to a
pure classical van-der-Waals gas.  As the latter, symmetric and neutron matter
develop first order phase transitions. This is no longer the case for
multi-component systems as asymmetric nuclear matter where protons and neutrons
occupy in momentum space different Fermi spheres.  Within the relativistic
mean-field model used in \cite{Muel:1995} quantal multi-component systems
change their state of aggregation by second order phase transitions. Hence, if
a phase transition is going to happen in a charge asymmetric nuclear system
signals will be washed out, complicating the identification of critical
phenomena. Clearly, this is a strict result only for equilibrium thermodynamics
based on consideration of infinite matter. Transport-like non-equilibrium
conditions as encountered in a heavy ion collision underly their own special
dynamical and statistical laws as e.g. in fragmentation reactions
\cite{Gaitanos:2008uc,Gaitanos:2011fy,Bondorf:1995ua}. Interestingly, as
discussed in those papers grand canonical thermodynamics is a successful
concept also for fragmentation processes.

The investigations in \cite{Muel:1995} are based on the non-linear extension of
the original Walecka model, allowing for self-interactions of the
scalar-isoscalar  field. Those approaches are completely phenomenological in
nature without attempting to relate the model parameters to an underlying
theory. Here, we follow a different approach by using the DDRH nuclear field
theory providing an easy to handle and flexible description of equilibrated 
cold and hot nuclear matter. DDRH theory, developed some time ago in Giessen,
\cite{Le:95,Fuchs:95,HoKeLe,JonLen,Le:2004}, is a density dependent field
theory with interactions derived by Dirac-Brueckner theory from free space
$NN$-interactions. The one-boson exchange picture on which DDRH theory is
based, gives access to the full spectrum of scalar, pseudo-scalar, and
vector interaction channels, typically described in terms of meson-exchange
interactions. The medium dependence is taken into account by vertex functionals
depending on Lorentz-invariant binomials of the baryon field operators. 
Their mean-field expectation values are evaluated by relations using the 
self-consistent scheme of Dirac-Brueckner Hartree-Fock theory. In this sense,
the density dependent relativistic hadron (DDRH) field theory is a 
parameter-free \emph{ab initio} description of nuclear matter. In particular,
the microscopic ansatz gives access to details  unreachable for 
phenomenological approaches. One of those regions is the contribution of
isovector-scalar interactions as realized in nature by the $a_0(980)$ meson.
The importance of the corresponding scalar-isovector mean-field for asymmetric
nuclear matter was pointed out in our previous work \cite{JonLen, Le:2004},
deriving for the first time the density dependence of the $\delta$-$NN$ vertex
by DBHF calculations. Since then, the interaction channel is being included as
a standard tool in many nuclear interaction studies, see e.g. 
\cite{Gaitanos200424, Dalen2013}.  Obviously, this is a short-range phenomenon
acting in competition with the vector meson repulsion.  However, the 
consequences are quite different and important: scalar fields modify the 
relativistic effective masses of baryons. Hence, they change the mechanical
properties. Since isovector-scalar fields lead in asymmetric nuclear matter
to counterbalanced modifications of proton and neutron effective masses we have
to expect important modifications of the dynamical and, especially, the 
thermodynamical laws.  Here, we present for the first time extensions of DDRH
theory to equilibrated nuclear matter at finite temperature. To a large extent,
the results obtained in the present work are representative for any type of
nuclear field theory based on density dependent meson-baron vertices. 
However, an essential and unique
feature over other work is the inclusion of scalar-isovector interactions. In a
relativistic approach the corresponding self-energies are leading to a splitting
of the proton and neutron relativistic effective masses in asymmetric matter,
thus changing the inertial properties of the particle species in an asymmetric
manner. An important built-in property of DDRH theory is the conservation of the
Hugenholtz-van Hove (HVH) theorem \cite{Hug:58}. Rearrangement self-energies,
accounting for the static polarization of the medium, are playing the essential
role \cite{Fuchs:95} in conserving the HVH relation. Since thermodynamical
consistency is fulfilled at all densities and temperatures, we are avoiding
artificial effects from violations of the HVH theorem which will appear
inevitably when density dependent effective interactions like a Brueckner
G-matrix are used but rearrangement self-energies are neglected.

The nuclear many-body theoretical background is discussed in
sect.\ref{sec:DDRH}, the relevant thermodynamical quantities are introduced in
sect.\ref{sec:Thermo}. In sect. \ref{sec:SymNucMat} we present our results on
the EOS in mean-field approximation of symmetric nuclear matter and in sect.
\ref{sec:AsymNucMat} those for asymmetric nuclear matter.  We also compare the
DDRH results to those obtained from the DD-ME2 model \cite{Lala:2005}, which is
a phenomenological derivative of our DDRH theory. In this model the density
dependent vertices are fitted directly to nulear data. Typically, DD-ME models
neglect the $\delta$ mean-field.  Additionally, in order to emphasize the
impact of the density dependent formulation on various properties of nuclear
matter, the results from the NL3\cite{Lala:1997}, the QHD model model with
constant interaction vertices \cite{Serot:1979} and the non-relativistic Skyrme
type model \cite{Chabanat1997710} are shown. For this purpose, we use the QHD
model with $\sigma$, $\omega$ and $\rho$ mesons in its original formulation
\cite{Serot:1979, Serot:1986p2564}.  In particular, we discuss the influence of
the isovector-scalar self-energies on the binodal and spinodal structures.
Conclusions are drawn and an outlook is given to open questions and future work
in sect.\ref{sec:Summary}.
%
%
\section{Density Dependent Hadron Field Theory}\label{sec:DDRH}

\subsection{The DDRH Lagrangian and Energy Momentum Tensor}\label{ssec:Lagrangian}

An important step forward in understanding the saturation properties of infinite
nuclear matter was achieved by theories describing in-medium interactions
microscopically. Using relativistic Dirac-Brueckner (DB) it was found that the
pertinent problem of non-relativistic G-matrix calculations, namely by always
ending up at the Coester-line and missing the empirical saturation point of
infinite nuclear matter, could be overcome. With standard free-space
$NN$-interactions, reproducing well the $NN$ scattering observables, the
empirical saturation properties of nuclear matter could be described
convincingly well \cite{AnaCel, HorSer, TerMal,BroMac,BoeMal, HubWeb, JonLen}.
Using realistic nucleon-nucleon meson-exchange potentials, in-medium
interactions are derived by complete resummation of (two-body) ladder diagrams.
Since full-scale DB calculations for finite heavy nuclei are not feasible a
practical approach is to derive from infinite matter DB results an equivalent
energy density functional. The Kohn-Sham \cite{Koh} and the Hohenberg-Kohn
\cite{HoKo} theorems confirm the existence of a general density functional,
although they do not provide a guideline for construction. Based on the work in
\cite{BroTok, BoeMal} we have derived in \cite{Le:95, Fuchs:95} a fully
covariant and thermodynamically consistent field theory by treating the
interaction vertices as Lorentz-scalar functionals of the nucleonic field
operators. The DDRH approach, unlike common relativistic mean field (RMF)
models, accounts for quantal fluctuations of the baryon fields even in the
ground state.  In \cite{HoKeLe} the DDRH theory has been applied to asymmetric
nuclear matter and nuclei far from stability. An appropriate set of coupling
constants has been derived, now including isoscalar ($\sigma$,$\om$) and
isovector ($\delta$,$\rho$) vertices. In turn, a phenomenological approach to
DDRH theory was proposed by Typel and Wolter \cite{TyWo}, trying to derive the
density dependence of the vertices empirically by fits to nuclear data.  Since
then, a large variety of purely phenomenological models has been formulated by
several groups and are being applied successfully to nuclei over almost the full
mass table.

In the fully microscopic \emph{ab initio} approach of the DDRH
theory one considers nuclear systems composed of protons and neutrons, described 
by Dirac field operators $\Psi_q$ with $q=p,n$. Interactions are derived in
the one boson exchange approximation by using a set of mesons, mainly
acting as virtual fields and thus providing the interactions among the nucleons. 
Our relativistic boson exchange model includes
pseudoscalar ($\pi,\eta$), scalar ($\sigma,\delta/a_0(980)$) and
vector mesons ($\omega,\rho$). The Lagrangian is given by
\begin{equation}\label{eq:Lagrange}
\mathcal{L}=\mathcal{L}_N+\mathcal{L}_M+\mathcal{L}_\mathrm{int}.
\end{equation}
The fermionic part for the matter fields $\Psi_q$ is of standard Dirac-form:
\begin{equation}\label{eq:Dirac}
\mathcal{L}_N=\sum_{q=p,n}{\overline{\Psi}_q\left(i\gamma_\mu\partial^\mu-M_q
\right)\Psi_q} \quad .
\end{equation}
In the meson sector of the theory we have to distinguish the scalar 
($\alpha=\sigma,\delta/a_0(980)$) and pseudoscalar ($\alpha=\eta,\pi$) mesons,
obeying equations of the type
\begin{equation}\label{eq:mesons1}
\mathcal{L}^{(\alpha)}_M=\frac{1}{2}\left(\partial_\mu\varphi_\alpha\partial^\mu\varphi_\alpha-m^2_\alpha\varphi^2_\alpha\right)\quad .
\end{equation}
The vector mesons ($v=\omega,\rho$) are subject to
\begin{equation}\label{eq:mesons2}
	\mathcal{L}^{(v)}_M=-\frac{1}{2}\left(F^{(v)}_{\mu\nu}F^{(v)\mu\nu}-m^2_vA^{(v)}_\mu
A^{(v)\mu}\right) \quad ,
\end{equation}
with the field strength tensors of the vector mesons
\begin{equation}
	F^{(\omega)}_{\mu\nu}=\partial_\mu A^{(\omega)}_\nu-\partial_\nu A^{(\omega)}_\mu, \quad \vc{F}^{(\rho)}_{\mu\nu}=\partial_\mu \vc{A}^{(\rho)}_\nu-\partial_\nu \vc{A}^{(\rho)}_\mu -\hat{\Gamma}_\rho(\hat{\rho})\left( \vc{A}^{(\rho)}_\mu\times \vc{A}^{(\rho)}_\nu\ \right), 
\end{equation}
where the $\rho$-meson field-tensor includes the non-Abelian isospin term 
assuming the coupling to the conserved isovector current. Note, however,
that the self-interaction term does not contribute in mean field approximation

Of special interest is the interaction part
$\mathcal{L}_\mathrm{int}=\mathcal{L}_{ps}+\mathcal{L}_s+\mathcal{L}_v$ which includes
the pseudoscalar vertices
\begin{equation}\label{eq:Lapscal}
\mathcal{L}_{ps}=\Gh_{\eta}(\hr)\overline{\Psi}\gamma_5\Psi\varphi_{\eta}+\frac{f_{\eta}}{m_{\eta}}\overline{\Psi}\gamma_5\gamma_\mu\Psi\partial^\mu\varphi_{ps}
+\Gh_{\pi}(\hr)\overline{\Psi}\gamma_5\vc{\tau}\Psi\vc{\varphi}_{\pi}+\frac{f_{\pi}}{m_{\pi}}\overline{\Psi}\gamma_5\gamma_\mu\vc{\tau}\Psi\partial^\mu\vc{\varphi}_{\pi}
\end{equation}
the scalar vertices,
\begin{equation}\label{eq:Lascal}
\mathcal{L}_{s}=\Gh_s(\hr)\overline{\Psi}\Psi\varphi_s+\Gh_\delta(\hr)\overline{\Psi}\vc{\tau}\Psi\vc{\varphi}_\delta
\end{equation}
and the vector meson interactions, respectively,
\begin{equation}\label{eq:Lavec}
\mathcal{L}_{v}=\Gh_\omega(\hr)\left(\overline{\Psi}\gamma_\mu\Psi
A^{(\omega)\mu}+\frac{f_\omega}{m_\omega}\overline{\Psi}\sigma_{\mu\nu}\Psi
F^{(\omega)\mu\nu}\right)+\Gh_\rho(\hr)\left(\overline{\Psi}\vc{\tau}\gamma_\mu\Psi\vc{A}^{(\rho)\mu}
+\frac{f_\rho}{m_\rho}\overline{\Psi}\sigma_{\mu\nu}\vc{\tau}\Psi
\vc{F}^{(\rho)\mu\nu} \right) \quad .
\end{equation}
A Lagrangian of the same type, but with bare coupling constants
$\Gamma_\alpha(\rho)\to g_\alpha=const.$, is used for free space $NN$
scattering, serving to fix the bare coupling constants.  The full set of mesons
with the respective coupling constants is used to obtain first the free space
T-matrix and the in-medium Dirac-Brueckner G-matrix. As discussed in
\cite{JonLen} the Driac-Brueckner Hartree Fock calculations are performed in
asymmetric nuclear matter, making it possible to derive the full set of isoscalar and
isovector self-energies. The self-energies contain via u-channel processes also
pion and eta-meson contributions.  They are in fact important for a proper
description of the density dependence of the self-energies.  As discussed in
refs. \cite{Fuchs:95, JonLen, HoKeLe} the self-energies are used to derive
effective density dependent Hartree-vertices.  Because, owing to parity
pseudo-scalar mesons do not develop classical, condensed fields neither the pion
nor the eta-meson fields contribute to the mean-field sector of the theory. They
are, however, present in dynamical processes leading to excitations of the
system. Here, we are considering nuclear thermodynamics only on the mean-field
level, thus leaving investigations of dynamical fluctuations to a future study.
Hence, for the present work only the scalar and the vector interaction
Lagrangians, Eq. \eqref{eq:Lascal} and Eq. \eqref{eq:Lavec}, respectively, will
be of direct relevance once we have derived the DBHF vertices.

The Lagrangian of Eq.~\eqref{eq:Lagrange} is highly non-linear in the fermionic
field operators contained in the density operator $\hr$. At the
theoretical-mathematical level, this is a wanted property because it allows us
to keep full control over all changes of the systems and imposes covariance of
the field equations.
The derivation of the vertex functionals and their explicit determination in
mean-field approximation as functions of the density is discussed in
\cite{HoKeLe}.

With the standard Legendre-transformation we obtain the DDRH energy momentum
tensor $T^{\mu\nu}$ \cite{Fuchs:95,Le:2004}. Of particular interest is the
Hamiltonian density
\beq
\mathcal{H}\equiv T^{00}&=&\pb\gamma_0\Sigma^{0(0)}\p+\pb \left(M-\Sigma^{s(0)}\right)\p+\sum_{\alpha=\sigma,\delta,\pi}\left((\partial^0\phi_\alpha)^2-\oover{2}\left[\partial_\lambda\phi_\al\partial^\lambda\phi_\al-m_\al^2\phi^2_\al\right]\right)\crcr
&&+\sum_{\al=\om,\rho}\left(\partial^0A_\lambda F^{(\al)\lambda0}-\oover{2}\left[m^2_\al A_\mu^{(\al)}A^{\mu(\al)}-F_{\lambda\rho}^{(\al)}F^{\lambda\rho(\al)}\right]\right)
\label{eq:T00}
\eeq
The full energy momentum tensor $T^{\mu\nu}$ includes rearrangement
contributions induced by the functional derivatives of the vertex functional
\cite{Le:95,Fuchs:95,HoKeLe}, modifying in particular the pressure density
$P\sim T^{ii}$.

By the standard techniques of finite temperature quantum many-body theory
\cite{FetterWalecka}, we introduce hadronic chemical potentials $\mu_q$ and the
corresponding number operators $N_q$. Covariant thermodynamics will be discussed
below. Here, we only indicate the connection to the more familiar
non-relativistic formulation. This is achieved by a canonical transformation
\be
K=H-\sum_\eta{\mu_\eta N_\eta}
\ee
from which the grand partition function is obtained
\be
Z_G=e^{-\beta\Omega}=\Tr\, e^{-\beta K}
\ee
and the statistical operator
\be
\rho_G=Z^{-1}_Ge^{-\beta\Omega}=e^{-\beta(\Omega-K)}
\ee
is serving to perform the thermodynamical average of observables over the grand
canonical ensemble. Explicit expression are derived below.

\subsection{The DDRH Field Equations and Mean-Field Approximation}\label{ssec:FieldEq}
From the above Lagrangian we derive by variation with respect to the
Dirac-adjoint fermion field operators $\overline{\Psi}$ wave equations of
Dirac-type,
\begin{equation}
\left(\gamma_\mu(i\partial^\mu-\Sigma^\mu_v)-M-\Sigma_s\right)\Psi=0
\quad .
\end{equation}
They include the vector and scalar self-energies $\Sigma^\mu_v$ and
$\Sigma_s$, respectively \cite{Le:95}.  The variation
leads to two distinct types of self-energies,
\begin{eqnarray}
\Sigma&=&\Sigma^{(b)}+\Sigma^{(r)} \label{eq:SelfE}\\
\Sigma^{(b)}&=&\diff{\Ld_{int}}{\pb}\\
\Sigma^{(r)}&=&\diff{\Ld_{int}}{\hr}\frac{\delta\hr}{\delta\pb}\label{eq:rearr}
\end{eqnarray}
The bare self-energies $\Sigma^{(b)}$ are of the conventional structure as also
obtained in a theory with density-independent, constant vertices. Since our
vertex functionals are derived from Dirac-Brueckner theory, they correspond to
the standard DBHF self-energies \cite{terH:87,JonLen}.
The rearrangement self-energies $\Sigma^{(r)}$ in Eq.~\eqref{eq:rearr} represent
an essential new feature of generic character for a density dependent formalism.
It is seen that these contributions originate from the variation of the vertex
functionals, thus describing the response of interaction vertex on a change of
the medium. Physically, $\Sigma^{(r)}$ accounts for the dynamical rearrangement
effects of the nuclear medium by virtue of polarization \cite{Le:95,Le:2004}.

The meson field equations are given by the standard Klein-Gordan
equations for the scalar and pseudo-scalar fields,
\begin{equation}
\left(\partial_\mu\partial^\mu+m^2_\alpha\right)\Phi_\alpha=\Gh_\alpha(\hat{\rho})\overline{\Psi}
\hat{O}_\alpha\Psi
\label{eq:KlGo}
\end{equation}
where $\hat{O}_\alpha \in \{1,\gamma_\mu,\gamma_5 \}\otimes\{1,\bf{\tau} \}$ and
derivative terms may contribute as well.  The vector fields obey the
Proca-equation
\begin{equation}
\partial_\mu F^{(\alpha)\mu\nu}+m^2_\alpha
A^\nu=\Gh_\alpha(\hr)\overline{\Psi}\hat{J}^{\nu}_\alpha\Psi
\label{eq:Pro}
\end{equation}
where $\pb \hat{J}^{\nu}_\alpha\Psi$ denotes the corresponding
vector current. The only difference is that the vertices depend now on
the background medium. For our purpose, the Bose-fields are far off
their mass shell. They are virtual fields which are fully determined
by their nucleonic sources, contributing only in $t-$ and
$u-$channel processes among nucleons.

The field equations are to be solved for a given nuclear matter ground state
configuration. Using the mean.field approximation we specify the expectation
values of the functionals. Let $\rho=\langle\hr \rangle$ be the ground state
expectation value of the operator $\hr=\hr(\bar{\Psi}_q, \Psi_q)$ which is a
Lorentz-scalar functional of the proton and neutron, respectively, field
operators $\Psi_q$. Using $\hr=\rho+\delta\rho$ we expand the vertex functionals
according to
\be\label{eq:VertSeries}
\Gamma_\alpha(\rho+\delta\rho)=\Gamma_\alpha(\rho)+\frac{\partial \Gamma_\alpha(\hr)}{\partial \hr}\Big|_{\rho}\delta \rho+\cdots
\ee
Taking the ground state expectation value of the vertex, the first order
correction $\delta\rho= \hr-\rho$ vanishes identically, The first non-vanishing
correction is given by
\be
C(\hr)=\langle \left(\hr-\rho \right)^2\rangle=\langle \hr\rangle^2-\rho^2 \quad ,
\ee
determined by the quantal fluctuations of $\hr$ with respect to the reference
value $\rho$. In ground state calculation we neglect terms of this and higher
order consistently. Therefore, in nuclear matter the meson field equations are
always determined by source terms with a strength given by
$\Gamma_\alpha(\rho)$, which is a number given as a  function of $\rho$.

Thus, neglecting consistently all non-stationary fluctuations around the
ground state expectation values, we obtain the mean-field equations
\begin{eqnarray}
\left(\vec{\nabla}^2-m^2_\alpha\right)\Phi_\alpha&=&-\Gamma_\alpha(\rho)\langle\pb\mathcal{O}_{(\alpha)}\psi\rangle\\
\left(\vec{\nabla}^2-m^2_\alpha\right)A^{(\alpha)\mu}&=&\Gamma_\alpha(\rho)\langle\pb\mathcal{O}^{(\alpha)\mu}\psi\rangle
\quad ,
\end{eqnarray}
where the fields are now classical scalar and vector fields, respectively. 

In the following we will apply the mean-field approximation to the full
Lagrangian, Eq. \eqref{eq:Lagrange}. As a consequence of time reversal
invariance and parity conservation, the pseudo-scalar and pseudo-vector fields
will become identical to zero in the ground state.  However, there is still an
implicit contribution of these fields to the of the remaining density dependent
couplings, since they are derived from the full Dirac-Brueckner theory.  This
can be seen with the help of the Fierz transformation and decomposition of the
exchange diagrams in terms of direct channels. \cite{JonLen}

As already indicated in the previous section, we will use the DDRH theory in
mean field approximation, with the explicit inclusion of the $\sigma$, $\delta$,
$\omega$ and $\rho$ meson mean-fields. The interaction part of the Lagrangian
then simplifies to 

\begin{equation}
	\mathcal{L}_\textrm{int} = \Gamma_\sigma(\rho)\bar{\psi}\Phi_\sigma\psi +\Gamma_\delta(\rho)\bar{\psi}\vc{\tau}\vc{\Phi}_\delta\psi-\Gamma_\omega(\rho)\bar{\psi}\gamma_\mu A^{(\omega)\mu}\psi - \Gamma_\rho(\rho)\bar{\psi}\gamma_\mu\vc{\tau}\vc{A}^{(\rho)\mu}. 
	\label{eq:LagrangeMFT}
\end{equation}

The first derivative terms in Eq.~\eqref{eq:VertSeries} lead to the
mean-field rearrangement self-energies, including scalar and vector
components:
\begin{eqnarray}
\Sigma^{s(r)}&=&\sum_\alpha{\left(\frac{\partial \Gamma_\alpha}{\partial
 \hr}\Big|_{\rho}\langle\pb\mathcal{O}_\alpha\psi\rangle\Phi_\alpha \right)}\\
\Sigma^{v(r)}&=&\sum_\alpha{\left(\frac{\partial \Gamma_\alpha}{\partial
 \hr}\Big|_{\rho}\langle\pb\mathcal{O}^{(\alpha)}_\mu\psi\rangle A^{(\alpha)\mu}
 \right)} \quad .
\end{eqnarray}

The self-energies are decomposed into isoscalar and isovector parts:
\begin{eqnarray}
\hat\Sigma^{s(0)}&=&\hat\G_\sigma(\hr)\phi_\sigma+\hat\G_\delta(\hr)\vc{\tau}\cdot\vc{\Phi}_\delta
\crcr &=&\hat\Sigma^{s(0)}_0+\vc{\tau}\hat{\vc{\Sigma}}_1^{s(0)}
\end{eqnarray}
and correspondingly, for the vector self-energies
\begin{eqnarray}
\hat\Sigma^{\mu(0)}&=&\hat\G(\hr)A^{(\om)\mu}+\hat\G_\rho(\hr)\vc{\tau}\cdot\vc{A}^{(\rho)\mu}\crcr
&=&\hat\Sigma_0^{(0)\mu}+\vc{\tau}\hat{\vc{\Sigma}}_1^{(0)\mu}.
\end{eqnarray}
where we have left out the photon field.
For infinite nuclear matter we find explicitly in Hartree
approximation \cite{HoKeLe} the non-vanishing self-energies
\begin{eqnarray}
\Sigma^{s(0)}_q(\hat\rho)&=&\G_\sigma(\hat\rho)\Phi_\sigma+\tau_q\G_\delta(\hat\rho)\Phi_\delta\crcr
\Sigma^{0(0)}_q(\hat\rho)&=&\G_\om(\rho)A_0^{(\om)}+\tau_q\G_\rho(\hat\rho)A^{(\rho)}_0\crcr
\Sigma^{0(r)}(\hat\rho)&=&\diff{\Gamma_\om}{\hat\rho}A_0^{(\om)}\rho_0+\diff{\Gamma_\rho}{\hr}A_0^{(\rho)}\rho_1
-\diff{\Gamma_\sigma}{\hr}\Phi_\sigma\rho^s-\diff{\Gamma_\delta}{\hr}\Phi_\delta\rho_1^s\quad
, \label{eq:SelfEn0}
\end{eqnarray}
where proton ($q=p$) and neutron ($q=n$) contributions are indicated and
$\tau_{q}=\pm1$ denotes the corresponding expectation values of the isospin
$\tau_3$ operator, respectively. All vertex derivatives are to be evaluated at
$\hr=\rho$. Since isospin symmetry requires that the vertices must depend only
on isoscalar quantities the rearrangement self-energies do not depend on the
nucleonic charge state. Defining the proton and neutron vector densities by
$\rho_q=\langle0|\pb_q\gamma_0\p_q|0\rangle$ and similarly the corresponding
scalar densities by $\rho^s_q=\langle0|\pb_q\p_q|0\rangle$ we find the isoscalar
($I=0$) and isovector ($I=1$) vector and scalar densities, respectively,
\begin{eqnarray}\label{eq:defDens}
\rho_I&=&\rho_n+(-)^I\rho_p\\
\rho^s_I&=&\rho^s_n+(-)^I\rho_p \quad .
\end{eqnarray}
It can be easily verified that in Eq.~\eqref{eq:SelfEn0} the isovector pieces
are in fact of quadratic order in $\rho_1$ and $\rho^s_1$,
respectively.

The proton and neutron wave functions, respectively, obey
Dirac equations of standard form
\begin{equation}
\left[\gamma_\mu\left(i\partial^\mu-\hat\Sigma^\mu_q\right)-\left(M_q-\hat{\Sigma}^s_q
\right)\right]\p_q=0,
\end{equation}
but with the self-energies as defined as in Eq.~\eqref{eq:SelfE} and
constructed by means of the results in Eq.~\eqref{eq:SelfEn0}.

Since we include interactions in the scalar-isovector channel represented by the
$\delta/a_0(980)$ meson the effective mass of the nucleons is
explicitly isospin dependent in DDRH theory \cite{JonLen,HoKeLe},
\begin{equation}
M^*_q=M-\G(\rho)\phi_\sigma-{\tau_q}\G_\delta(\rho)\phi_\delta.
\label{eq:effMass}
\end{equation}

%
\section{Thermodynamics of Nuclear Matter} \label{sec:Thermo}
\subsection{Covariant Thermodynamics}
In the covariant formulation of thermodynamics it is expedient to express all
equations in  terms of Lorentz scalars and Lorentz four-vectors.  For this
purpose we follow the steps from \cite{Furn:90} by introducing the thermal
potential $\alpha=\beta\mu$ and the thermal time-like four vector
$\beta_{\mu}=\beta u^\mu$, where
$u^{\mu}=\frac{1}{\sqrt{1-{v^{2}}/{c^{2}}}}\left(1,\vc{v}\right)$ is the fluid
velocity four-vector and $\beta={1}/{T}$ the inverse temperature. 

A system of a set of conserved currents $j^\mu_c$ in equilibrium is described by
the energy momentum tensor $T^{\mu\nu}$ and the entropy flux $\sigma^\mu$. These
quantities are connected by the first law of thermodynamics \cite{Kapusta:1989,
Serot:1986p2564}
\begin{align}
	\label{eq:cov-first-law-th}
	\beta_{\nu}\text{d}T^{\nu\mu}&=\td \sigma^{\mu}+ \sum_{c}\alpha_{c} \td j_{c}^{\mu}.
\end{align}
The pressure $P$ can be expressed in terms of the primary functions by
\begin{align}
\label{eq:pressure2}
P\beta^{\mu}&=-\beta_{\nu}T^{\nu\mu}+\sigma^{\mu}+\sum_{c}\alpha_{c}j^{\mu}_{c}.
\end{align}
In statistical quantum mechanics the thermodynamic functions are related to
ensemble averages of quantum-mechanical operators.  This is usually achieved by
defining a grand partition function $Z$ and a corresponding four-vector
potential $\Phi^{\mu}(\alpha_{c},\beta_{\nu})$. The partition function is
\begin{align}
\label{eq:cov-part-func}
Z&\equiv\Tr \left[\exp\left(\int\td \mathcal{F}_{\mu}\left[\sum_{c}\alpha_{c}j^{\mu}_{c}-\beta_{\nu}T^{\nu\mu}\right]\right)\right]
\end{align}
where $\mathcal{F}_{\mu}$ denotes the four-dimensional infinitesimal surface
element.  By varying  $\beta_{\nu}$ and $\alpha_{c}$ one can derive  a
differential equation, that connects $\Phi^{\mu}$ with the energy momentum
tensor and the conserved currents,
\begin{align}
\td \Phi^{\mu}=T^{\mu\nu}\td\beta_{\nu}-\sum_{c}j^{\mu}_{c}\td \alpha_{c},
\end{align}
providing the covariant thermodynamic laws
\begin{subequations}
		\label{eq:cov_therm_laws}
	\begin{align}
		\label{eq:cov_therm_laws_T}
		&T^{\mu\nu}=\left(\frac{\partial \Phi^{\mu}}{\partial \beta_{\nu}}\right)_{\alpha_{c}},\\
		\label{eq:cov_therm_laws_J}
		&j^{\mu}_{c'}=-\left(\frac{\partial \Phi^{\mu}}{\partial \alpha_{c'}}\right)_{\beta_{\mu},\alpha_{c},c \neq c'}.
	\end{align}
\end{subequations}
Furthermore, after differentiating Eq.~\eqref{eq:pressure2} and using the first
law of thermodynamics, Eq.~\eqref{eq:cov-first-law-th}, we arrive at the
covariant form of the Gibbs relation:
\begin{align}
	\Phi^{\mu}=-P\beta^{\mu}=\beta_{\nu}T^{\nu\mu}-\sigma^{\mu}-\sum_{c}\alpha_{c}j^{\mu}_{c}
\end{align}
In the comoving frame $u^{\mu}=(1,0,0,0)$, thus,  Eq.~\eqref{eq:cov-part-func} reduces to
\begin{align}
	\label{eq:Partition-Function}
	&Z\equiv \sum_{n}\langle n |e^{-\beta (\hat{H}- \sum_{c}\mu\hat{N}_{c})}| n\rangle=\Tr\; e^{-\beta (\hat{H}- \sum_{c}\mu_{c}\hat{N}_{c})},
\end{align}
where $\hat{H}$ is the Hamiltonian describing the system and $\hat{N}_{c}$ the
number operator. In Eq.~\eqref{eq:Partition-Function} the trace has to be taken
as a sum over all energy and particle eigenstates $|n \rangle$.
With this, the above expressions lead to a connection between the thermodynamic
potential and other thermodynamic functionals:

\begin{subequations}
	\begin{flalign}
		\Phi^{\mu}(\alpha_{c},\beta_{\nu})&= -\frac{1}{V}\ln Z u^{\mu} =\frac{\Omega(T,\mu,V)}{TV}u^{\mu}\\
		\rho_c &=-\frac{1}{V}\frac{\partial \Omega }{\partial \mu_c}\\
		\mathcal{S}&=\beta \left(\mathcal{E}-\Omega-\sum_b\mu_b \rho_b\right)
	\end{flalign}
\end{subequations}

According to the Hugenholtz-Van Hove theorem \cite{Hug:58}, the thermodynamic
and the hydrostatic pressure must coincide, i.e.
\begin{align}
	\label{eq:thermo-consist}
	P=\rho^2\frac{\partial \mathcal{F}}{\partial \rho}=\frac{1}{3}\sum_{i=1}^3\langle  T^{ii}\rangle,
\end{align}
with the free energy $\mathcal{F}=\mathcal{E}/\rho-T\mathcal{S}$. In
Eq.~\eqref{eq:thermo-consist} the thermal average is used, which for an
arbitrary operator $\hat{O}$ is given by the prescription:
\begin{align}
	\label{eq:thermal-expval}
	\langle \hat{O}\rangle= \frac{\Tr \left[e^{\beta(\hat H-\sum_{c}\mu_{c}\hat{N}_{c})}\hat O\right]}{\Tr \left[e^{\beta(\hat H-\sum_{c}\mu_{c}\hat{N}_{c})}\right]}= \Tr \left[\hat \rho \hat O \right].
\end{align}

Consider now the DDRH Hamilton operator, which in mean field approximation (MFA)
is given by
\begin{align}
  \hat{H}(\hat{\rho})=T^{00}&=\Big[\sum_b~\bar{\psi}_b\left(i\vc{\gamma}\pmb{\nabla}+\gamma_0\Sigma^{0(0)}_b(\hat\rho)+\left(M-\Sigma^S_b(\hat{\rho}\right)\right)\psi_b \nonumber\\
    &+\frac{1}{2}m_\sigma^2\phi_\sigma^2+\frac{1}{2}m_\delta^2\phi_\delta^2 - \frac{1}{2}m_\omega^2 {A_0^{(\omega)}}^2-\frac{1}{2}m_\rho^2 {A_0^{(\rho)}}^2\Big],
	\label{eq:DDRH-Hamiltonian}
\end{align}
where the sum over $b$ includes all baryons considered in the model. An
important consequence of the  MFA is the cancellation of the rearrangement term
in the Hamiltonian.  However, since the interaction vertices are functionals of
the density operator $\hat{\rho}$ the calculation of the partition function
\eqref{eq:Partition-Function} needs to be examined carefully.  To derive $Z$ it
is necessary to evaluate a trace over the eigenstates in Fock space.  This is
only possible if the exponential function can be decomposed into a sum of
independent terms.  To fulfill this requirement in the DDRH model, the
Hamiltonian should be approximated by a one-body operator.  For this purpose the
density-dependent interaction vertices $\hat{\Gamma}_{b\alpha}$ are expanded
around the thermal equilibrium density mean value $\rho_0=\langle
\hat{\rho}\rangle$, keeping only terms up to the first order in
$(\hat\rho-\rho_0)$ (see also Eq.~\eqref{eq:VertSeries})
\begin{align}
  \nonumber
	\Gamma_\alpha(\hat{\rho})=\Gamma_\alpha(\rho_0)+\left.\frac{\partial \Gamma_\alpha}{\partial\hat\rho}\right|_{\hat\rho=\rho_0}(\hat\rho-\rho_0)+\mathcal{O}\left((\hat\rho-\rho_0)^2\right).
\end{align}
With this assumption the expectation values of the vertex functionals become
functions of the density, i.e.\ $\langle\hat{\Gamma}(\hat\rho)
\rangle\rightarrow\Gamma(\rho_0)$. Hence, the Hartree-vertices are now
implicitly temperature dependent, $\Gamma(\rho,T)=\Gamma(\rho(T))$.  On the
other hand this approach implies an expansion for the Hamilton operator
\begin{align}
  \nonumber
  H(\hat{\rho})=H^0(\rho_0)+H^R( \hat{\rho} )
\end{align}
with
\begin{align}
  \nonumber
	H^R( \hat{\rho} )=  \Sigma^{(r)}(\rho_0) \left( \hat{\rho}-\rho_0 \right) =\sum_b
  \Bigg(&
  \frac{\partial\Gamma_{b\sigma}}{\partial \hat{\rho}}\Bigg|_{\hat{\rho}=\rho_0}\phi_\sigma\rho_b^s
  +\frac{\partial\Gamma_{b\delta}}{\partial \hat{\rho}}\Bigg|_{\hat{\rho}=\rho_0}\phi_\delta \tau^3_b \rho_b^s\\
  +&\frac{\partial\Gamma_{b\omega}}{\partial \hat{\rho}}\Bigg|_{\hat{\rho}=\rho_0}A^0_\omega \rho_b
  +\frac{\partial\Gamma_{b\rho}}{\partial \hat{\rho}}\Bigg|_{\hat{\rho}=\rho_0}A^0_\rho \tau^3_b\rho_b   \Bigg) \left( \hat{\rho}-\rho_0 \right) \end{align}
This shows that the in-medium correlations described by the density-dependent
couplings indeed are inducing a rearrangement perturbation in the Hamilton
operator.  Given this approximation, all operators in the exponential of the
partition function are now diagonal. This makes an exact calculation feasible.
From the stationary solution of the Dirac equation in MFA we find the energies
for  baryons and anti-baryons at the equilibrium nuclear matter density $\rho_0$
\begin{align}
	\varepsilon^\pm_b(k)=\pm E^*_b(k) + \Sigma^{0(0)}_b+\Sigma^{(r)}(\rho_0),
  \label{}
\end{align}
with  $E^*_b(k)=\sqrt{k^2+M^{*2}_b}$. Considering the normal ordered products of
the baryon fields  we can now go through the usual steps for the calculation of
the partition function. Indeed, introducing effective baryon masses and
effective chemical potentials as
\begin{align}
	M^*_b&\equiv M_b-\Sigma_b^S\nonumber\\
	\nu_b&\equiv \mu-\Sigma_b^0=\mu-\Sigma_b^{0(0)}-\Sigma^{(r)},
\end{align}
leads to the same expression for the baryonic part of the thermodynamic
potential as in the non-interacting case. This time, however, the baryon masses
and the chemical potentials are replaced by their effective values. Since the
meson fields are treated as classical fields their contribution to the partition
function is trivial.
Thus we can split $\Phi^\mu$ in a baryon, mean field and rearrangement part
\begin{align}
	\label{eq:Phi-DDRH}
	\Phi^\mu= \sum_b\Phi^\mu_b+ \Phi^\mu_{\textrm{MF}}+\Phi^\mu_\textrm{R}
\end{align}
with
\begin{align}
	\Phi^\mu_b &= -\sum_s\int\frac{\mathrm{d}^3k}{(2\pi)^3}\left[ \ln\left(1+e^{-\beta(E^*_k-\nu_b)}\right)+\ln\left( 1+e^{-\beta(E^*_k+\nu_b)} \right) \right]u^\mu\\
	\Phi^\mu_{\textrm{MF}} &=\beta\left( \frac{1}{2}m_\sigma^2\phi_\sigma^2+\frac{1}{2}m_\delta^2\phi_\delta^2 - \frac{1}{2}m_\omega^2 {A_0^{(\omega)}}^2-\frac{1}{2}m_\rho^2 {A_0^{(\rho)}}^2\right)u^\mu\\
	\Phi^\mu_{R}&= -\beta\rho_0\Sigma_0^{(r)} u^\mu.
\end{align}

The fluctuations around the equilibrium density coming from density dependent
correlations show up as rearrangement terms in  $\Phi_\textrm{R}^\mu$.
Consequently, the pressure $P=-\Phi^0/\beta^0$ is also modified by additional 
rearrangement contributions, which are crucial to fulfill the 
Hugenholtz-Van Hove theorem.  
However, the rearrangement parts cancel out in the entropy and energy density. 
For the latter for example this can be verified by applying 
Eq.~\eqref{eq:cov_therm_laws_T} to $\Phi^\mu$,
\begin{align}
	\mathcal{E}=\!\!\frac{\partial \Phi^0}{\partial \beta^0}\Bigg|_{\alpha_b}\!\!\!\!\!\!\!\!=\sum_b\left( 2\!\!\int\!\!\frac{\td^3{k}}{(2\pi)^3}E^*_k\left( f_B(\nu_b)+\bar{f}_b(\nu_b) \right)+\Sigma^{0(0)}_b\rho_b\right) +\frac{1}{2}\left(\sum_{s=\sigma,\delta}\!\!\!\! m^2_s\Phi^2_s -\sum_{v=\omega,\rho} \!\!\!\! m^2_v A_0^{v2}\right)
	\label{eq:Energy-Density}
\end{align}
Note, that in the calculation of the above derivative all parameters
$\alpha_b=\beta\mu_b$ have to be held fixed.

In thermal equilibrium the meson fields should be chosen such that they minimize
$\Phi^\mu$, 
i.e.\ $\left(\frac{\partial\Omega}{\partial\chi}\right)_{\beta,\mu_b}=0$ 
for $\chi\in\{\phi_\sigma,\phi_\delta,A_0^\omega, A_0^\rho\}$.
This results in the following equations
\begin{subequations}
\begin{align}
	\phi_{\sigma} & =  \sum_b \frac{{\Gamma}_{b\sigma}}{m^2_\sigma}~\rho^{s\phantom{(3)}}_b \equiv  \sum_b \frac{{\Gamma}_{b\sigma}}{m^2_\sigma}\phantom{\tau^3_b}\cdot2\int\!\!\!\!\frac{\td^3{k}}{(2\pi)^3}~\frac{M^*_b}{E^*_b}\left( f_B(\nu_b,T)+ \bar{f}_B(\nu_b,T) \right) \\
  \phi_{\delta} & =  \sum_b \frac{{\Gamma}_{b\delta}}{m^2_\delta}~\rho^{s(3)}_b  \equiv   \sum_b \frac{{\Gamma}_{b\delta}}{m^2_\delta}\cdot2\tau^3_b\int\!\!\!\!\frac{\td^3{k}}{(2\pi)^3}~\frac{M^*_b}{E^*_b}\left( f_b(\nu_B,T)+\bar{f}_B(\nu_b,T) \right) \\
	A_0^{\omega} & =  \sum_b \frac{{\Gamma}_{b\omega}}{m^2_\omega}~\rho_b^{\phantom{(3)}}   \equiv  \sum_b \frac{{\Gamma}_{b\omega}}{m^2_\omega}\phantom{\tau^3_b}\cdot2\int\!\!\!\!\frac{\td^3{k}}{(2\pi)^3}~\left( f_B(\nu_b,T)- \bar{f}_B(\nu_b,T) \right) \\
	A_0^{\rho}   & =  \sum_b \frac{{\Gamma}_{b\rho}}{m^2_\rho}~\rho^{3\phantom{(3)}}_b   \equiv  \sum_b \frac{{\Gamma}_{b\rho}}{m^2_\rho}\cdot2 \tau^3_b\int\!\!\!\!\frac{\td^3{k}}{(2\pi)^3}~\left( f_B(\nu_b,T)- \bar{f}_B(\nu_b,T) \right)
  \label{}
\end{align}
\end{subequations}

In the limit of $T\rightarrow0$ the particle distribution function becomes the
well known step function, $f_B(\nu_B)\rightarrow\Theta\left( \nu_b-E^*_k
\right)\equiv\Theta(k_{F_b}-k)$, while the anti-particle distribution function
$\bar{f}_B$ vanishes completely. This defines the Fermi momenta and Fermi
energies of the baryons as $\nu_b=E^*_{F_b}=\sqrt{k_{F_b}^2+M^{*2}_b}$.
The solution of the baryon densities and, accordingly, the vector meson fields
becomes trivial for vanishing $T$,
while the scalar densities can be found by a self-consistent solution of the
transcendental equation
\begin{align}
	\rho^s_b=\frac{M^*_b}{2\pi^2}\left( E^*_{F_b} k_{F_b}-M^{*2}_{F_b}\ln\left[ \frac{E^*_{F_b}+k_{F_b}}{M^*_b} \right] \right)
\end{align}
The energy density and pressure of cold nuclear matter are then given by:
\begin{align}
	\mathcal{E}(T=0)=\sum_b\frac{1}{4}\left( 3 E_{F_b} \rho_b+M^*_b\rho^s_b\right)+\sum_b\frac{1}{2}\left( \rho_b\Sigma^{0(0)_b}+\rho^s_b\Sigma^{S}_b \right)\\
	P(T=0)=\frac{1}{4}\sum_b\left( E_{F_b}\rho_b-M^{*}_b\rho^s_b  \right)+\frac{1}{2}\sum_b\left( \rho_b\Sigma_b^{0(0)}-\rho^s_b\Sigma^S_b \right)+\rho\Sigma^{(r)}
	\label{eq:EoS-DDRH-T0},
\end{align}
where $\rho=\sum_b\rho_b$ is the total baryon density.

\subsection{The Density and Temperature Dependence of DDRH Vertices and Self-Energies}
Typically, nuclear matter mean-field models assume tacitly that interactions are
essentially unaffected by temperature. Already quite early, ter Haar and
Malfliet \cite{terH:87} have addressed that question by means of finite
temperature DBHF calculations. They indeed find a weak temperature dependence of
the resulting G-matrix interaction. A closer inspection of the DB-equations at
$T>0$ reveals the reason for that behavior.

\begin{figure}[ht]
	\begin{center}
		\includegraphics{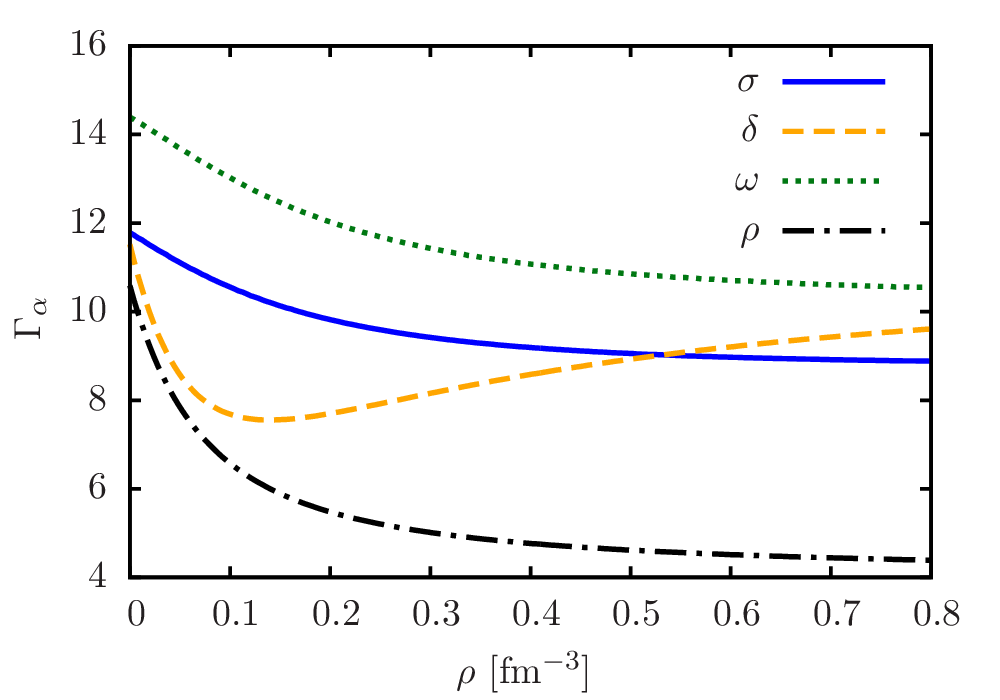}
		\caption{(Color online) Density dependence of the DDRH nucleon-meson 
			vertices at $T=0$ \cite{HoKeLe}.}
		\label{fig:G_DDRH}
	\end{center}
\end{figure}

\begin{figure}[ht]
	\begin{center}
	\includegraphics{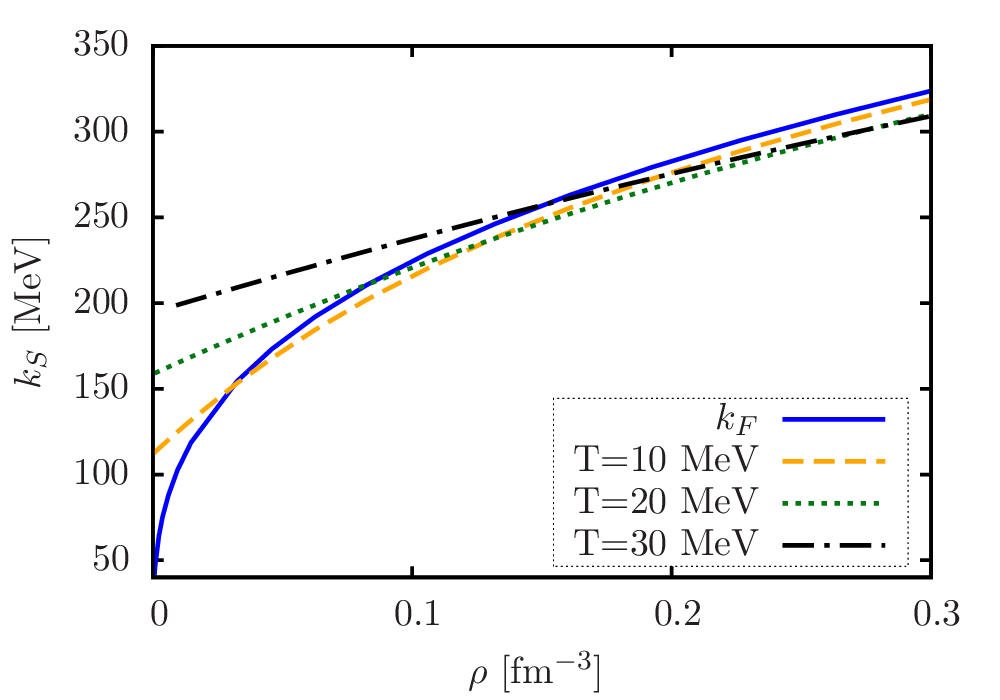}
	\end{center}
	\caption{(Color online) Comparison of $k_S$ versus the Fermi momentum $k_F$ for $T$=10, 20 and 30 MeV.}
	\label{fig:kS_vs_kF}
\end{figure}

\begin{figure}[ht]
	\includegraphics{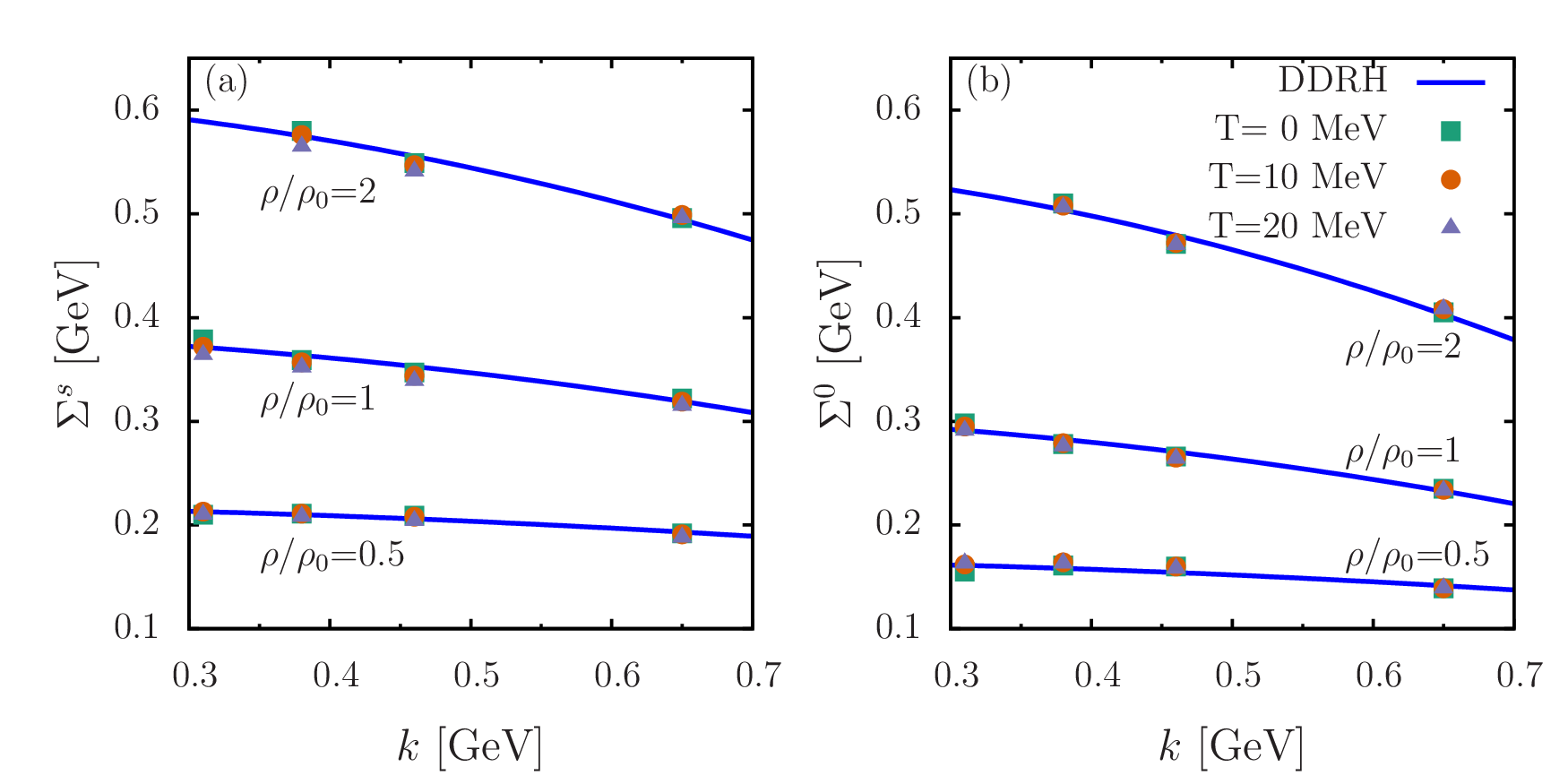}
	\caption{(Color online) The real part of the scalar, (a), and vector, (b), Dirac-Brueckner
		self-energies at different temperatures and densities,
		\protect\cite{terH:87}, compared to the quadratic approximation discussed in
	the text.}
	\label{fig:DB-Sigma-k}
\end{figure}

The medium and temperature dependence of the G-matrix is introduced by  two
sources, namely the Pauli blocking of the Fermi-sphere of occupied single
particle states and the baryon self-energies. In leading order contributions
from the Pauli principle, i.e. Fermi-statistics, are dominant. 
Therefore, preparatory to the discussion of the equation of state, we should
take a closer look at the DDRH vertices first. 
In Fig.~\ref{fig:G_DDRH} the density-dependence of the DDRH vertices is shown.
One can nicely see that the density-dependence has its most impact in the lower
density region, and becomes less significant for $\rho>0.5 \mathrm{ fm}^{-3}$.
Basically, this circumstance arises from the fact that in the high density
region the contributions from intrinsic particle fluctuations are small compared
to the ones coming from the occupied Fermi sea states.
The functional behavior of the $\delta$-meson vertex differs considerably from
the one of the other mesons. While all other vertices constantly fall off with
rising $\rho$ the $\delta$-meson vertex shows a special functional  behavior
with a minimum at $\rho\approx 0.14 \mathrm{~fm}^{-3}$. The inclusion of this
channel is a key feature of the DDRH model, which leads to a splitting of the
effective nucleon masses in asymmetric nuclear matter, as will be are
later.
The DDRH vertices are deduced from  comparing  the DDRH potential energy
with the Brueckner calculations \cite{Fuchs:95,HoKeLe}.
In general, the proper {DB} self-energies, $\Sigma^{\textrm{DB}}$, are momentum
dependent, while the DDRH self-energies are not, since they are calculated in
the mean-field approximation.
The usual approach to map $\Sigma^\textrm{DDRH}(\rho,T)$ on
$\Sigma^\textrm{DB}(k,\rho,T)$, is to calculate the average of $\Sigma^{\textrm{DB}}$
over the Fermi sea.
For the vector self-energies this implies
\begin{align}
	\Sigma^{\textrm{DDRH}}_\alpha(\rho,T) = \rho\frac{\Gamma^2_\alpha}{m^2_\alpha}=\frac{4}{(2\pi)^3}\frac{1}{\rho} \int\td^3k~\Sigma^\textrm{DB}_\alpha(k,\rho,T)\left( f_B - \bar{f}_B \right).
	\label{eq:Sigma-DDRH-DB}
\end{align}
For scalar self-energies $\rho$ has to be replaced by $\rho_s$ on the left hand
side of \eqref{eq:Sigma-DDRH-DB}.\\
In \cite{JonLen} it was shown that the {DB} self-energies can be approximated by
quadratic functions in the momentum $k$ in the vicinity of the Fermi momentum
$k_F$. Hence, an expansion of $\Sigma^\text{DB}$ up to the first order in $k^2$
around the Fermi momentum is necessary and sufficient to reproduce the {DB}
equation of state properly,
\begin{align}
	\Sigma^{\text{DB}}(k,\rho)\approx\Sigma_F^{\text{DB}}(\rho)+\left( k^2-k_F^2 \right) \Sigma'^{\textrm{DB}}_F(\rho),
	\label{eq:SDB-expansion}
\end{align}
with the definitions: $\Sigma^{\text{DB}}_F=\Sigma^\text{DB}(k_F,\rho)$,
$\Sigma'^{\text{DB}}_F=\frac{\partial\Sigma^\text{DB}(k,\rho)}{\partial k}\Bigg|_{k=k_F}$.\\
The Fermi momentum $k_F$ is given by the relation $\sqrt{k_F^2+M^{*2}}=\nu$ at
$T=0$. Thus, $k_F$ is not well defined for $T>0$ and should be replaced by a
more suitable quantity at finite $T$.
One possibility is to define the momentum $k_S$, where the Fermi distribution is
half its value at $k=0$,
$$
	f_B\left( k_S,T,\nu \right)=\frac{1}{2}f_B\left( 0,T,\nu \right)
$$
Since the weight-factor $k^2 f_B(k,T,\nu)$ in the momentum integral peaks near
$k_S$, it is useful to evaluate $\Sigma^\text{DB}$ around this momentum value
instead. Because $k_S\rightarrow k_F$ for $T\rightarrow 0$, this definition is
compatible to zero temperature calculations. In Fig.~\ref{fig:kS_vs_kF} $k_S$ is
plotted as a function of $\rho$ for some temperature values. One can see that
there is only a significant difference at low densities and high temperatures
between $k_S$ and $k_F$. The differences between $\Sigma^\text{DB}(k_S)$ and
$\Sigma^\text{DB}(k_F)$ are very small in this region. This was also found by
ter Haar and Malfliet in \cite{terH:86}.
Nevertheless, we will substitute $k_F$ by $k_S$ in \eqref{eq:SDB-expansion} from
now on to provide a well  defined expansion scheme for all temperatures.
In the range of moderate temperatures the {DB} self-energies show a very small
temperature dependence \cite{Sammarruca:2010p4681, terH:86}.
As an example, Fig.~\ref{fig:DB-Sigma-k} illustrates the temperature dependence
of $\Sigma^\text{DB}$ as obtained by \cite{terH:86}.
From this we conclude, that it is not only possible to apply the same quadratic
expansion of $\Sigma^{\text{DB}}$ as suggested by \cite{JonLen} at $T>0$, but
even to neglect the temperature dependence of the self-energies in a first order
approximation.
Substituting $\Sigma^{\text{DB}}$  by \eqref{eq:SDB-expansion} in
\eqref{eq:Sigma-DDRH-DB} gives the following expression for the momentum
corrected DDRH vertices:
\begin{align}
	\Gamma^2_\alpha &= \frac{4 m^2_\alpha}{(2\pi)^3}\frac{1}{\rho_{\alpha}}\Sigma_S\left( 1 +\frac{\Sigma'_S}{\Sigma_S}\cdot\frac{1}{\rho} \int \td^3k \left( k^2-k_S^2 \right)(f_B-\bar{f}_B) \right)\nonumber\\
	&\equiv \Gamma^2_{(0)\alpha} \left( 1+C^\alpha_S I_M \right)
\label{eq:Gamma-MC}
\end{align}
In \eqref{eq:Gamma-MC} $\rho_\alpha$ corresponds to $\rho$ and $\rho_s$ for
vector and scalar mesons, respectively.  $\Gamma_{(0)\alpha}$ is the first order
approximation to the dressed vertex, including only Hartree contributions from
{DB} self-energies. For vector mesons $\Gamma_{(0)\alpha}$ is density dependent,
while for scalar mesons it additionally depends on the temperature, since
$\rho_s$ is a function of $T$ at fixed $\rho$. In this case we can further
separate $\Gamma_{(0)s}$
\begin{align}
	\Gamma_{(0)s}(\rho, T)=\tilde\Gamma_{(0)s}(\rho) \cdot\sqrt{\frac{\rho_s(T=0)}{\rho_s(T)}} \equiv\tilde\Gamma_{(0)s}(\rho) \cdot I_s(\rho,T)
	\label{eq:Gs0-separation}
\end{align}
Obviously, $\Gamma_{(0)s}(T\!=\!0)=\tilde{\Gamma}_{(0)s}$.
The function $I_s$ provides an additional temperature correction to the scalar
vertex. It is important to note here, that a modification of $\Gamma_s$ causes a
modification of $\rho_s$, which in turn implies a different value for
$\Gamma_s$. Therefore, $I_s(\rho,T)$ must be calculated numerically by solving
the self-consistency problem.\\

The second term in \eqref{eq:Gamma-MC} incorporates momentum corrections to the
vertex, where $C_S^\alpha=\Sigma^{\alpha'}_S/\Sigma^\alpha_S$ and the momentum
integral $I_M$ is given by
\[
	I_M(\rho,T)=\frac{2}{\pi^2\rho}\int\td k~k^2 \left( k^2-k_S^2(T) \right)\left( f_B(T)-\bar{f}_B(T) \right).
\]
As already mentioned, the parameter $C_S$ can be assumed as independent of
density and temperature. However, the momentum integral depends explicitly on
both, $\rho$ and $T$,
\[
	I_M=I_M(\rho,T)
\]
At $T=0$ the expression for $\Gamma^2_\alpha$ simplifies to
\begin{align}
	\Gamma^2_\alpha(\rho,T=0)=\tilde\Gamma_{(0)\alpha}^2\left[ 1+C^\alpha_S\frac{2}{\pi^2\rho}\left( -\frac{2}{15}k_F^5 \right) \right]=\tilde\Gamma_{(0)\alpha}^2\left[ 1 - \frac{2}{5} C^\alpha_S k_F^2  \right]=\tilde\Gamma^2_{(0)\alpha}\left[ 1-\kappa C^\alpha_S\rho^{2/3} \right],
	\label{eq:Gamma-MC-T0}
\end{align}
with $\kappa=\frac{2}{5}(\frac{3}{2}~\pi^2)^{\frac{2}{3}}$.\\
As thermal excitations become important with increasing $T$, the momentum
correction integral $I_M$ will considerably differ from
$\kappa\rho^{\frac{2}{3}}$.
The constant factor $\kappa C^\alpha_S$ however turns out to be very small
leading to a modification of the scalar and vector couplings by only $0.8\%$ and
$0.1\%$, respectively \cite{HoKeLe}.
Therefore, the temperature dependence of $\Gamma_\alpha$ induced by $I_M$
remains small in the temperature and density ranges relevant for this work.
For further discussion, we introduce the function $I^\alpha_C$,
\begin{align}
	I^{\alpha}_{C}(\rho,T)\equiv \left[  \frac{\left( 1+C_S^\alpha I_M(T) \right)}{\left( 1+C^\alpha_S I_M(T\!\!=\!\!0) \right)}\right]^{\frac{1}{2}},
\end{align}
which describes the temperature dependent correction factor on $\Gamma_\alpha$
arising from the momentum correction integral. This gives the following
separation ansatz for $\Gamma_\alpha$:
\begin{align}
	\Gamma^\alpha=\Gamma^\alpha(T\!=\!0)\cdot I_C^\alpha(T)I_s^\alpha(T).
\end{align}
Note, that $I_s=1$ for vector mesons.

\begin{figure}[ht]
	\begin{center}
	\includegraphics{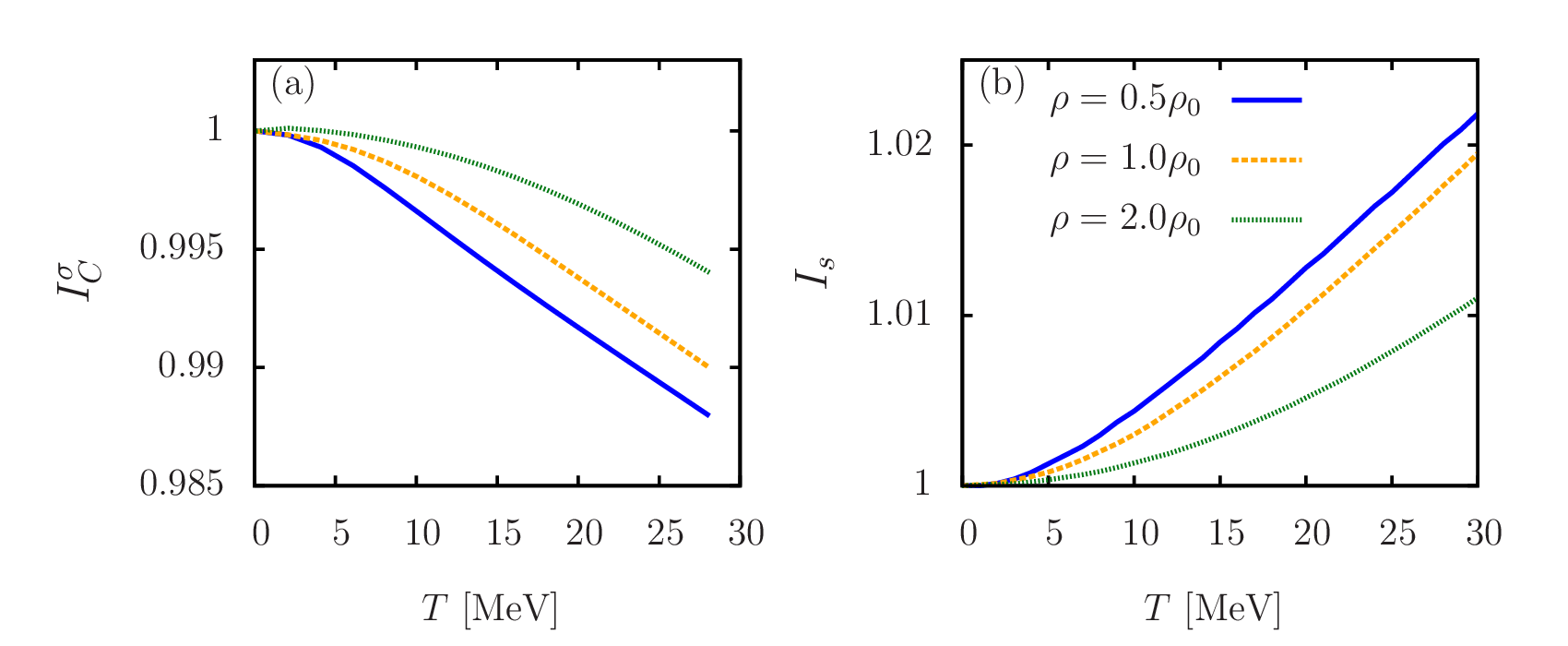}
	\end{center}
	\caption{(Color online) Temperature dependence of the scalar coupling
	functions $I_C^\sigma$ (a) and $I_s^\sigma$ (b) at different densities:
	$\rho=0.5\rho_0$ (blue line), $\rho=\rho_0$(orange dashed line), $\rho=2\rho_0$ (green dotted), where
	$\rho_0$ is the saturation density at $T=0$. Lower curves
	belong to lower densities.}
	\label{fig:Gs-T-rho}
\end{figure}

In Fig.~\ref{fig:Gs-T-rho} the  functions $I^\sigma_C(T)$ and $I^\sigma_s(T)$
for the $\sigma$-meson coupling are shown at various densities $\rho$. First of
all we see, that the temperature dependence becomes less significant with
increasing $\rho$. As expected, the momentum correction integral results in
small deviations of $\Gamma_\sigma$ from its zero temperature value. While
$I^\sigma_C$ falls off continuously with rising $T$ the scalar correction
function $I^\sigma_s$ shows an opposite behavior. Thus, the two effects
partially compensate each other.
In the range of moderate temperatures, where the liquid-gas phase transition
takes place, the net correction is  less than $0.5\%$ and it stays small even at
higher $T$.
For the vector coupling $\Gamma_\omega$ we find that the corrections are even
less then $0.2\%$. 

This analysis therefore shows, that the approximation 
\begin{align}
	\Gamma^{\textmd{DDRH}}(\rho, T) \approx \Gamma^{\textmd{DDRH}}(\rho, T=0)
\end{align} 
can be applied for the DDRH vertices.  The temperature dependent 
corrections would have nearly negligible small effects on the finite
temperature equation of state for temperatures below 100 MeV.  Besides, there
are only few experimental data available for warm nuclear matter and it comes
with large uncertainties as in the case of liquid-gas critical temperature, for
example \cite{Karnaukhov:2008p4861}. Hence, we will use the effective vertices
in first order approximation, assuming a density dependence only.  Within this
approximation our calculations are in very good agreement with other models, as
will be shown below. The temperature dependent momentum corrections can be used
in future calculations to fine-tune the equation of state properties, whenever
more precise data will be available.
\begin{figure}[ht]
\begin{center}
\includegraphics[scale=1]{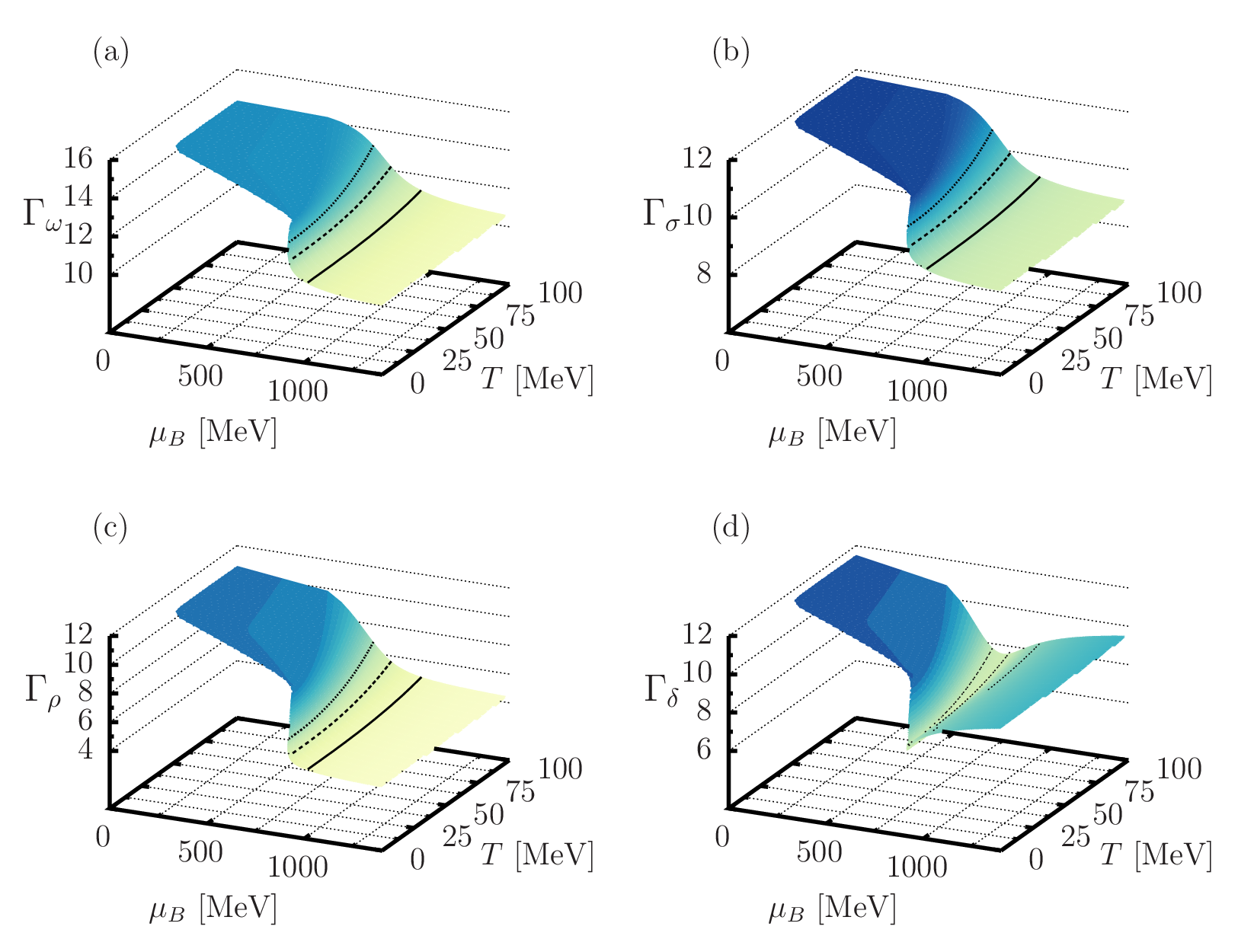}
\end{center}
\caption{(Color online) Dependence of the Hartree mean-field vertices on
chemical potential and temperature in symmetric nuclear matter. Results for the
vector vertices ($\omega,\rho$) and the scalar vertices ($\sigma,\delta$),
respectively, are displayed in the left and the right column, respectively. 
The dashed iso-lines indicate the constant density values 
of $0.5\rho_0$, $\rho_0$ and $2\rho_0$.}
\label{fig:fig2}
\end{figure}

In the temperature range considered here, $T<100$~MeV, the Fermi-Dirac
statistics is exerting only little or, at best, a moderate influence on the
G-matrix. Most part of the high energy tails of the thermal distributions are
suppressed by the vertex form factors typically used to regularize the
Bethe-Salpeter correlation integral.  These expected properties are well
reflected by our vertex functionals. The dependence of the vertices on chemical
potential and temperature in Hartree approximation are illustrated in
Fig.~\ref{fig:fig2}. The variations observed for high temperatures at chemical
potentials close to the nucleon mass, are corresponding to the increased
polarizability of low density nuclear matter. At these densities, the tails of
the thermal distributions are not yet suppressed by the vertex form factors,
hence leading to a stronger influence on the in-medium interactions. Note, that
the iso-lines in Fig.~\ref{fig:fig2} correspond to constant nuclear matter
densities.

In other studies the density-dependence of the meson-baryon vertices has been
determined in a phenomenological approach, where the functional form of the
vertices is adjusted to fit measured properties of symmetric and asymmetric
nuclear matter and spherical nuclei. The first phenomenological description 
was introduced by Typel and Wolter \cite{TyWo}. The parameters were further  
adjusted to calculations of giant multipole resonances  in \cite{Nik:2002}  
and \cite{Lala:2005}. Thus, in the present work we compare the results of 
the DDRH model to those obtained with the phenomenological DD-ME2 
parametrization of the vertices as given in \cite{Lala:2005}. 
In order to better understand the effects arising from density depentent 
effective couplings, we also show some results for the linear Walecka model with 
constant couplings, which we refer to as QHD. 

\section{Equation of State for Isospin Symmetric Nuclear Matter}
\label{sec:SymNucMat}

To begin with, we discuss our results for isospin symmetric nuclear matter at
$T>0$. An important quantity in the analysis of nuclear many-body properties is
the binding energy per nucleon,
\begin{align}
	\label{eq:BindingEnergy}
	E_\mathcal{B}=\frac{\mathcal{E}}{\rho}-M,
\end{align}
with the total baryon density $\rho=\sum_b\rho_b$ and $M=\sum_b\frac{\rho_b}{\rho}M_b$.
For the analysis of the different many-body effects it is helpful to separate
$E_\mathcal{B}$ in a kinetic and a potential part,
i.e. $E_\mathcal{B}=E_\mathrm{kin}+E_\mathrm{pot}$.
By inserting $\mathcal{E}$ from Eq.~\ref{eq:Energy-Density} in the above
equation and expressing the meson fields with the help of the corresponding
self-energies, we can write
\begin{align}
	E_\mathrm{kin}&=\frac{2}{\rho}\sum_{b}\left(\int\frac{\td^3 k}{(2\pi)^3}E^*\left( f_B+\bar{f}_B \right)-M^*_b\rho_b\right)\\
	E_\mathrm{pot}&=\frac{1}{2\rho}\sum_b \left( \Sigma^{0(0)}_{b}\rho_b -\Sigma^S_{b}\rho^s_b \right).
\end{align}

\begin{figure}[ht]
	\begin{center}
		\includegraphics{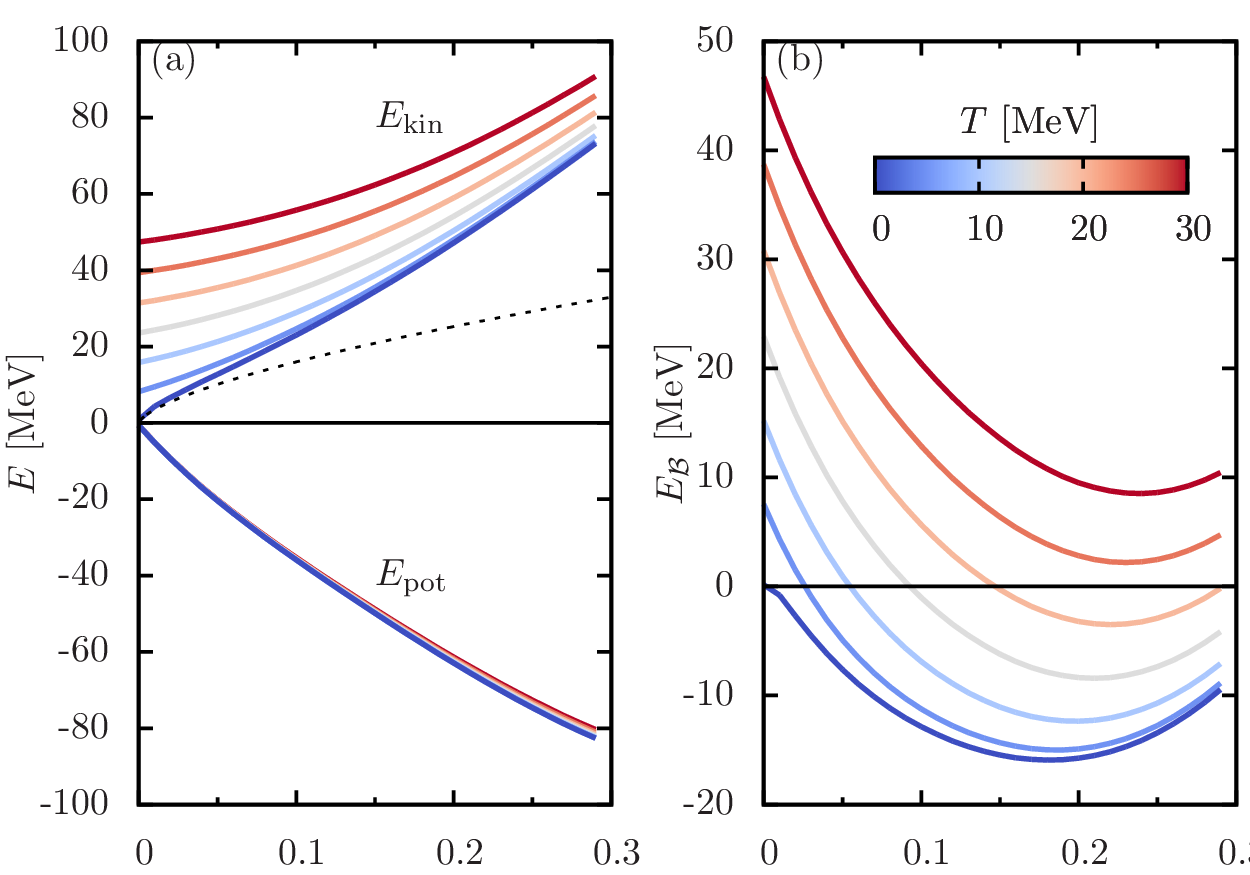}
	\end{center}
	\caption{(Color online) (a) Density dependence of the kinetic (upper curves),
		potential (lower curves). (b) Binding energy per particle within
		the DDRH approach. Isotherms for temperatures T=0 to T=30 MeV in
		equidistant steps of 5 MeV are displayed. Note, that lower curves correspond
		to lower temperatures. }
	\label{fig:BindingEnergy}
\end{figure}

In Fig.~\ref{fig:BindingEnergy} the results for the kinetic, potential and total
binding energy per particle in the DDRH model are given as functions of the
density for temperatures from 0 to 30 MeV.
Obviously, thermal excitations affect mostly the kinetic part of the energy,
while the potential energy is nearly unaffected by temperature effects.
The kinetic energy is also modified by many-body interactions through the
inclusion of the self-consistent effective masses $M^*_B$. At $T=0$ this can be seen by
comparing the DDRH result and the corresponding result for a
non-interacting gas. The latter is indicated by the dashed curve in
Fig.~\ref{fig:BindingEnergy}.
Because the effective mass is decreasing with density, nucleon-nucleon
interactions result in a repulsive effect in the kinetic energy.

The repulsion of $E_\mathrm{kin}$ is partially compensated by the potential
energy.  $E_\mathrm{pot}$ is negative in the whole density range and decreases
constantly with $\rho$. At lower temperatures the attraction of the potential
energy is strong enough to create a binding in the total energy.  The interplay
between the repulsive and attractive character of the two energies shows up as a
local minimum in $E_\mathcal{B}$, defining the saturation point $\rho_0$ of
nuclear matter.
In the DDRH calculation at $T=0$  the saturation point is found at
$\rho_0=0.18$ with $E_\mathcal{B}(\rho_0)=-15.94$, which is very close to {DBHF}
calculations. A comparison of the equilibrium properties at $T=0$ between the different 
models is given in Table \ref{tab:eqpropertiesT0}. 
With rising temperature the repulsion of $E_\mathrm{kin}$ becomes
stronger leading to a less bound system. For  $T>22$ MeV $E_\mathcal{B}$ stays
positive in the whole density range.
In the $\rho\rightarrow 0$ limit the kinetic energy, and thus $E_\mathcal{B}$,
comes very close to the classical value of an ideal gas, $E_\mathcal{B}=3/2 T$.

\begin{table}
	\caption{\label{tab:eqpropertiesT0} Comparison of the nuclear matter equilibrium properties at $T=0$.}
	\begin{ruledtabular}
		\begin{tabular}{lccccc}
			Model &  $\rho_0$ [fm$^-3$] & $E_\mathcal{B}(\rho_0)$ [MeV] & $K$ & $M^*_b$ & $E_\textrm{sym}$ \\
			\hline
		 DDRH & 0.181 & -15.94  & 282  & 518 & 26.7\\
		 DD-ME2 & 0.153 & -16.54 & 251 & 534& 32.3\\
		 NL3 \cite{Lala:1997} & 0.148 & -16.3 & 272 & 562 & 37.4 \\
		 SLy230a \cite{Chabanat1997710} & 0.160 & -16.0 & 230 & 655 & 32.0 \\
		 QHD & 0.192 & -15.4 & 530 & 524&33.7
	 \end{tabular}
 \end{ruledtabular}
\end{table}

\begin{figure}[h]
  \centering
\includegraphics{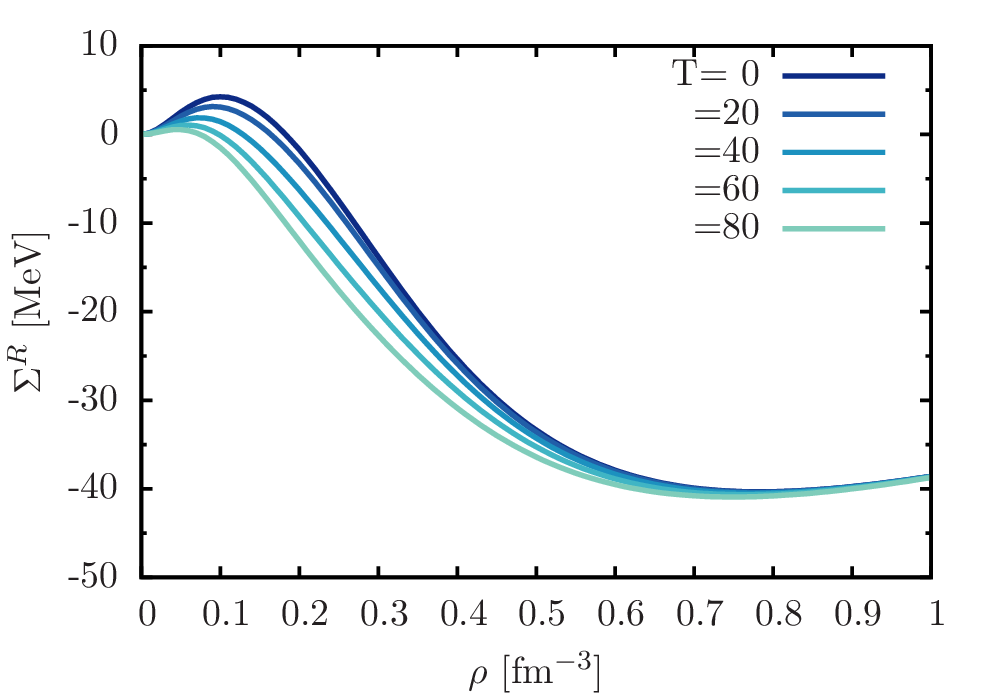}
\caption{(Color online) The rearrangement self-energy $\Sigma^{R}$ as a function
of baryon density $\rho$ at various values of temperature $T$ (in MeV). Lower
curves correspond to higher temperatures.}
\label{fig:SigmaR}
\end{figure}

The effect of density-dependent interactions on the equation of state in the
DDRH model can be better understood by examining the functional behavior of
the rearrangement self-energy $\Sigma^{R}$.
In Fig.~\ref{fig:SigmaR}  we show $\Sigma^{R}$  as a function of the baryon
density $\rho$ for symmetric nuclear matter at fixed temperatures.  Obviously,
$\Sigma^{R}$ is  very small compared to the Hartree self-energies
$\Sigma^{0(0)}$ ( $\approx 350$ MeV at saturation density $\rho_0$).  This
justifies the expansion of the Hamiltonian up to the first order in the density
deviation as described in the previous section.  The temperature dependence of
$\Sigma^{R}$ comes mainly from the scalar densities $\rho^s_b$ which saturate
for  $\rho\gg\rho_0$.  Therefore, $\Sigma^{R}$ becomes independent of
temperature in the high density limit.  In the region of nuclear matter
saturation density, $\rho_0$, the rearrangement contributions are more sensitive
to temperature changes.  However, the functional deviations are still rather
small here and one can consider $\Sigma^{R}$ as independent of temperature for
$T<20$ MeV.

\begin{figure}[h]
	\begin{minipage}{3.2in}\includegraphics{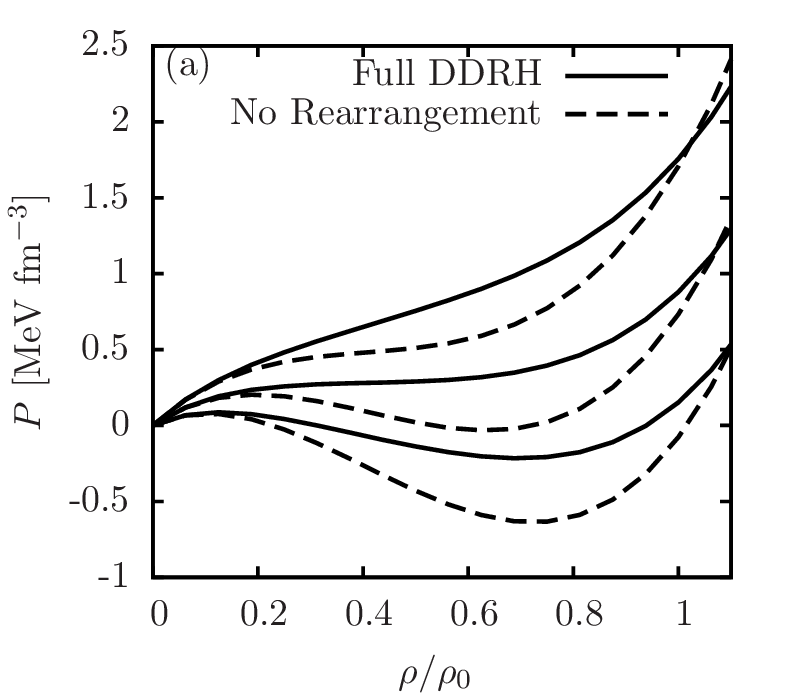}\end{minipage}\quad	\begin{minipage}{3.2in}\includegraphics{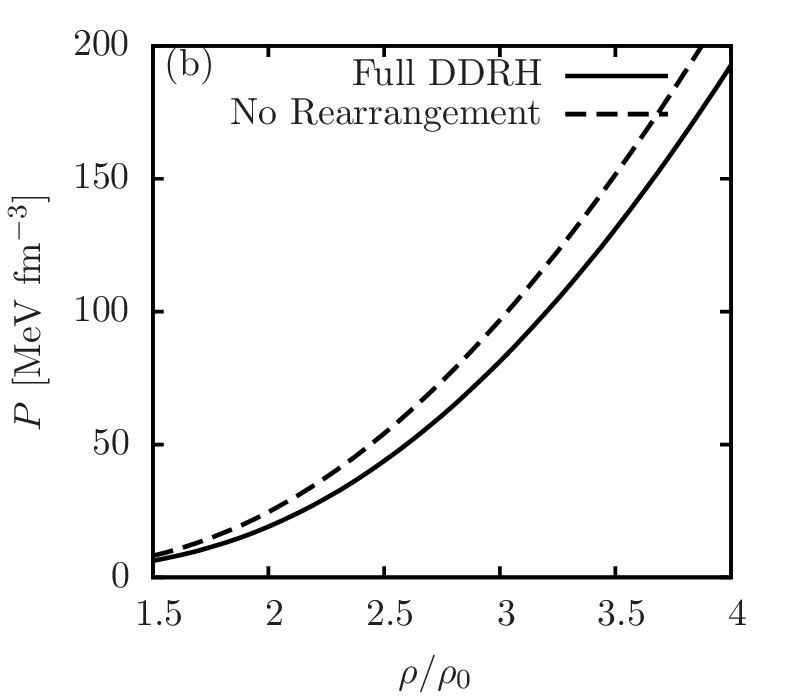}\end{minipage}
		\caption{The rearrangement contributions to the pressure. (a) Low density
		region for $T=10$, 15 and $20$ MeV (lower curves correspond to lower $T$).
	(b) High density region for $T=$20 MeV. }
	\label{fig:P-Rearr-05}
\end{figure}

The influence of the rearrangement energies on the pressure is illustrated in
Fig.~\ref{fig:P-Rearr-05}. The full DDRH calculation (solid lines) is compared
to calculations without rearrangement terms (dashed lines). In contrast to the
self-energies the rearrangement contributions have a significant effect on the
nuclear matter pressure.
Their modification on $P$ is even contrary in the low and high density region. While for $\rho<\rho_0$ the rearrangement terms cause an increase in the pressure density, their inclusion softens $P$ for $\rho>1.5\rho_0$. This is of course a direct consequence of the functional dependence of $\Sigma^{R}$ on $\rho$ and $T$, since $P^R=\rho\Sigma^{R}$.

\begin{figure}[h]
	\begin{minipage}{3.2 in} \includegraphics{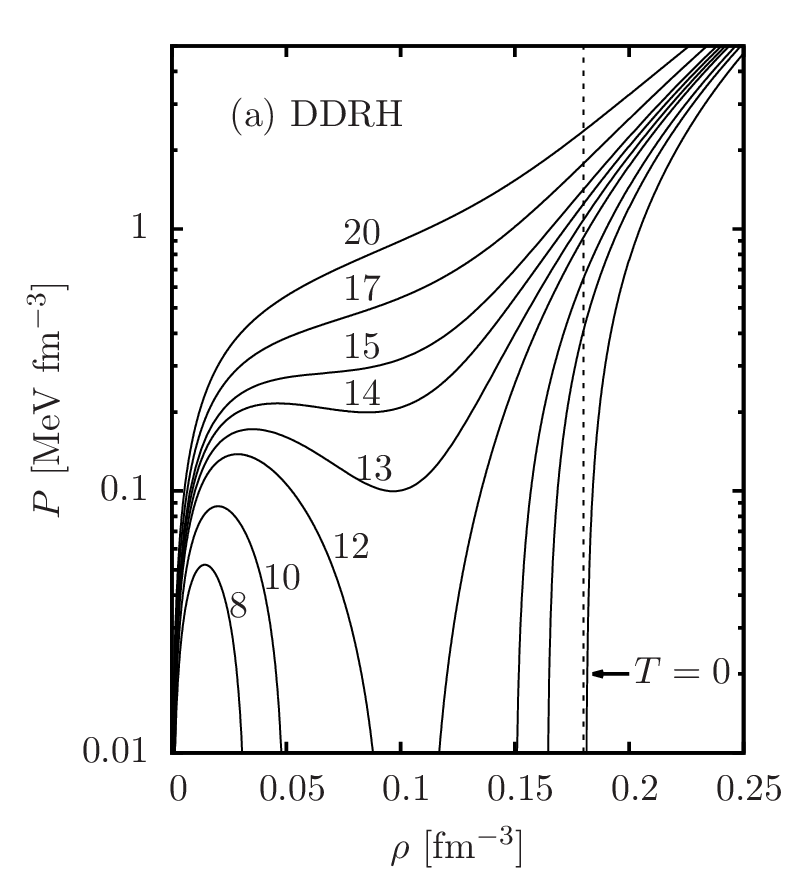}\end{minipage}\quad
		\begin{minipage}{3.5 in}\includegraphics{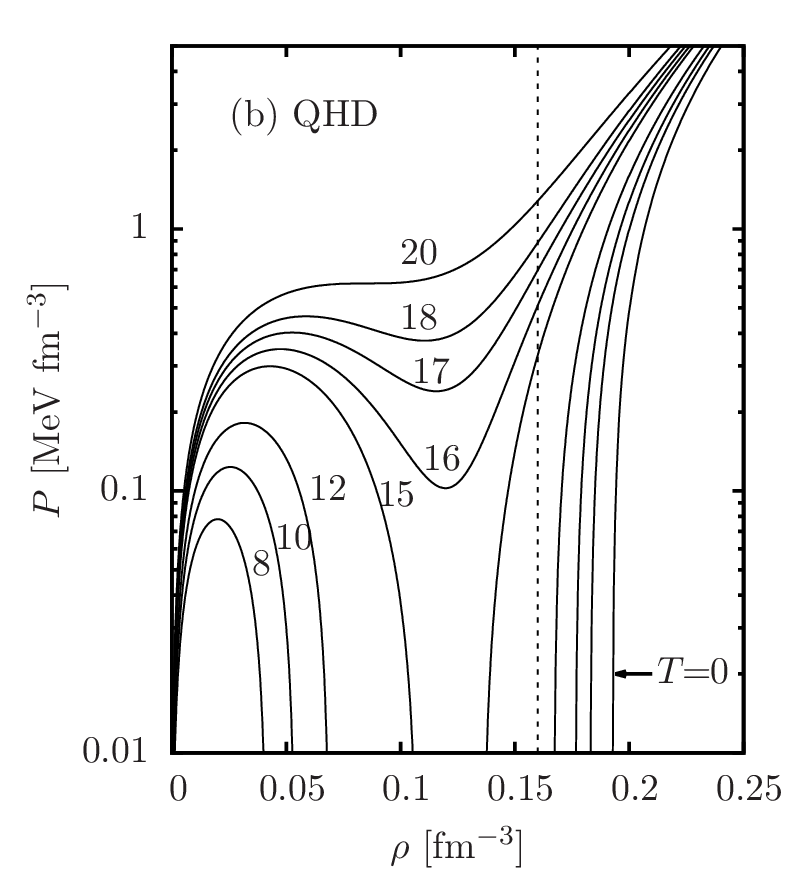}\end{minipage}
	\caption{Comparison of the equation of state $P(\varepsilon(\rho))=P(\rho)$ for symmetric nuclear matter at various temperatures. The numbers denote the temperature $T$ in MeV.}
			\label{fig:EoS-T-05}
\end{figure}

\begin{figure}[h]
\centering
\includegraphics{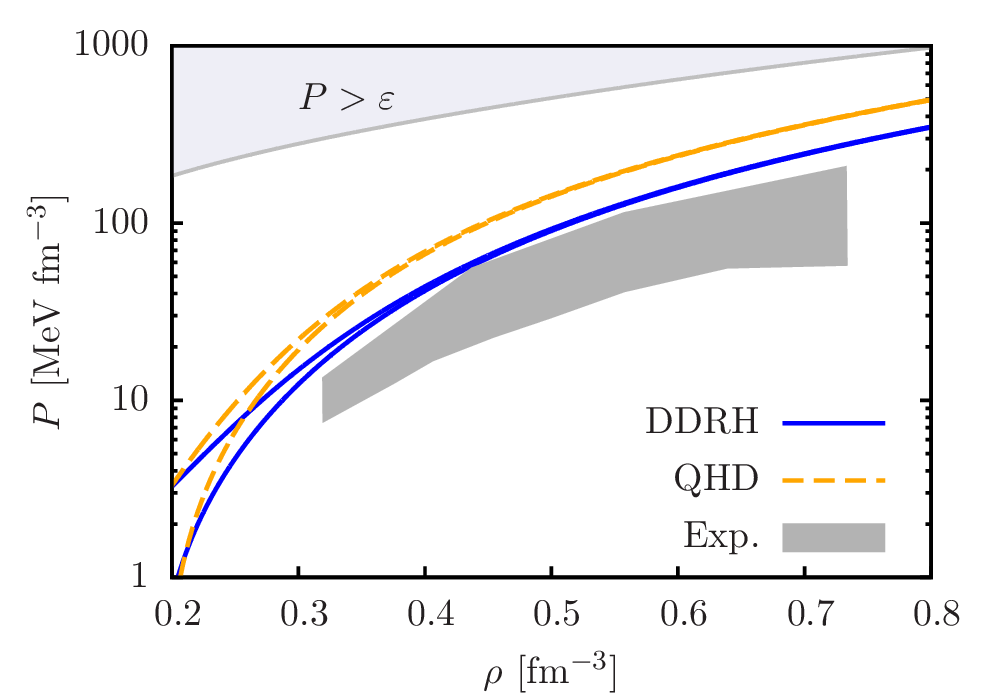}
\caption{(Color online) Functional behavior of the pressure $P$ in the high
density region for symmetric nuclear matter. The graph shows the results at
$T=0$ MeV (lower curves) and $T=20$ MeV (upper curves) for DDRH and QHD models.
The gray region corresponds to the region of pressures consistent with the
experimental flow data. The upper blue shaded area, labelled by $P>\varepsilon$
exhibits the causality limit.}
			\label{fig:EoS-T-05-highdens}
\end{figure}

The increase of the pressure at low densities plays a crucial role on the
liquid-gas phase transition region of nuclear matter.
Figure~\ref{fig:EoS-T-05} shows the pressure $P(T,\rho)$ of symmetric nuclear
matter as a function of the nucleon density $\rho$ for fixed values of the
temperature $T$ (isotherms).  The curves exhibit the characteristics of a
typical first-order liquid-gas phase transition.  In the low temperature region
the pressure first increases slightly with the density, then decreases to its
minimum point and finally returns to a continuous rising in the high density
region, where it asymptotically approaches the causality limit $P=\mathcal{E}$
(Fig.~\ref{fig:EoS-T-05-highdens}).
This functional behavior is very similar to the one of a classical van-der-Waals liquid \cite{Jaqaman:1983p4276}.

\begin{figure}[h]
	\centering
		\includegraphics{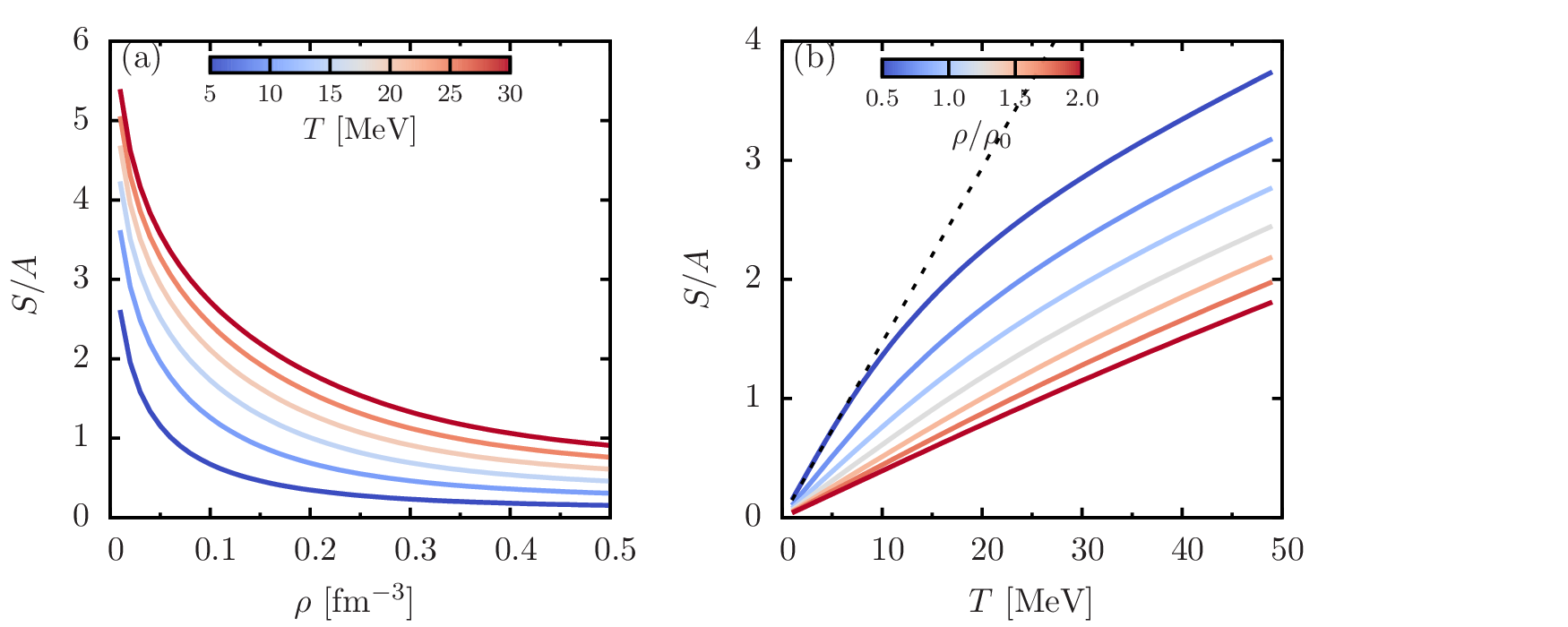}
	\caption{(Color online) Entropy per particle. Left: As a function of $\rho$
		for fixed $T$ from 5 to 30 MeV in steps of 5 MeV. The dashed line shows the
		classical limit at T=30MeV. Right: As a function of $T$ for fixed density
		values from 0.5$\rho_0$-2$\rho_0$ in steps of $\frac{1}{4}\rho_0$. The
	dashed line shows the limit of a Fermi liquid as explained in the text.}
	\label{fig:Entropy-SymNM}
\end{figure}

Next, we discuss the entropy in the DDRH approach. Entropy production plays an
important role in the determination of the mass fragment distribution in
multi-fragmentation events of heavy-ion collisions \cite{Peilert:1994p5036}.
In Fig.~\ref{fig:Entropy-SymNM} the density and temperature dependence of the
entropy per particle, $S/A=\mathcal{S}/{\rho}$, is shown. The left panel shows
$S/A$ as a function of density for fixed temperatures in the range of 0-30 MeV.
First of all we can see that the entropy increases with temperature and
significantly decreases with density.
This behavior is what is expected given the fact, that entropy is a measure of
thermal disorder.  At $\rho\rightarrow \infty$, $\mathcal{S}/\rho$ saturates at values in the range
of $0.5$-$1.0$ for the considered temperatures.  Additionally, there are two
notable limits. First, at low densities the entropy approaches the logarithmic
density dependence of a classical system, $S/A\sim-\ln(\rho)$.  This can be
ascribed to the fact that in the regime of very low densities as well as high
temperatures quantum effects become less important and thus the properties of a
classical system are recovered. On the other hand, the temperature dependence of
$S/A$ approaches a linear behavior at $T\rightarrow 0$, as shown on the right
panel of Fig.~\ref{fig:Entropy-SymNM}. This becomes even more pronounced  at
higher densities. In this regime, the system can be assumed as a Fermi liquid,
where the relation between $S$ is given by
\be
	\frac{\mathcal{S}_\mathrm{FL} }{A}=\frac{\pi^2}{2\rho}N_F T,
\ee
where
\be
N_F=\frac{E_Fk_F}{\pi^2}
\ee
is the density of states at the Fermi surface \cite{Rios:09}.
\begin{figure}[h]
	\centering
		\includegraphics{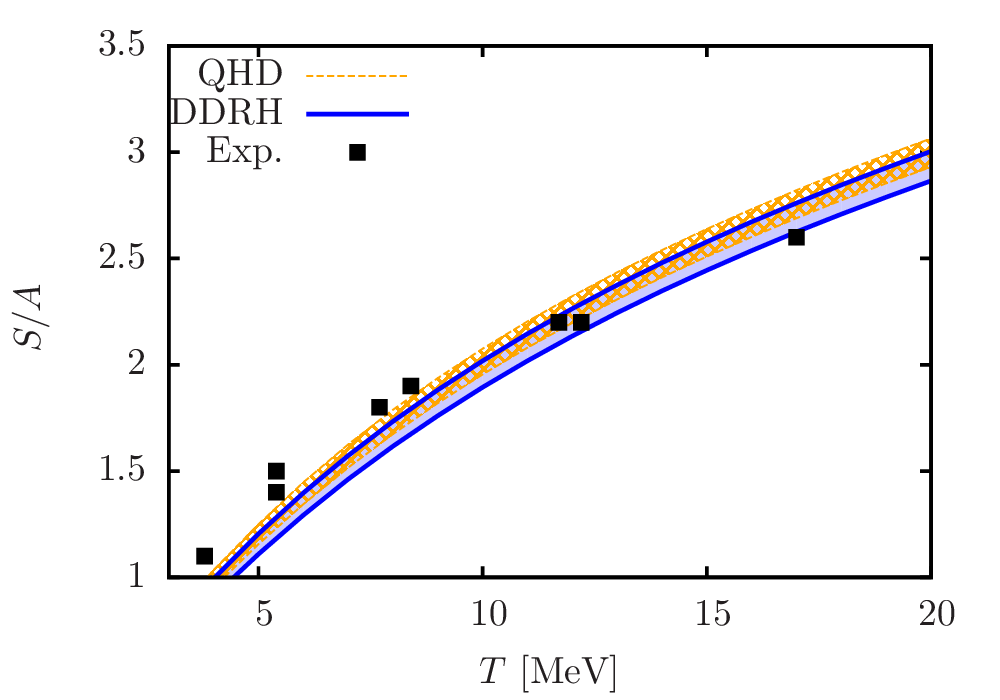}
		\caption{(Color online) Comparing the entropy per particle from the DDRH
		calculations to experimental results  \cite{Dzelalija:1995p5016}. The
		theoretical band corresponds to  $.047 \leq \rho\leq .053$ fm$^{-3}$,
		accounting for the uncertainties in the experimental analysis.}
	\label{fig:Entropy-SymNM-Compare}
\end{figure}

In Fig.~\ref{fig:Entropy-SymNM-Compare} we compare the DDRH calculation of
$S/A$ to results obtained from Au+Au collisions at energies between 100 and
400 AMeV \cite{Dzelalija:1995p5016}.
The experimental data was obtained from the study of the fragments which remain
after the collision. By assuming that the dense fireball created in the interior
of a collision is in thermal equilibrium, information on the entropy per
particle of the fireball can be obtained, in principle, from the fragmentation
remnants of the collision, see e.g. \cite{Bondorf:1995ua}.
However, this analysis comes with some uncertainties due to statistical
assumptions which might not be fully satisfied in a heavy ion collision. Apart
from that, one assumes that the freeze out density of the fireball is
$\rho\approx.3\rho_0$.
This provides a further uncertainty in the calculation.  To account for small
deviations from this central density we show the theoretical results in the
range of $.047\leq\rho\leq.053$ fm$^{-3}$.
Although, we find that in this density and temperature range the differences
between the DDRH and the simple QHD model are very small, the DDRH results
indicate a slightly better agreement with the data for temperatures higher than
10 MeV.
Nevertheless, one should note that a clear distinction between the models is
barely possible for this set of data.\\

\begin{figure}[h]
	\centering
		\includegraphics{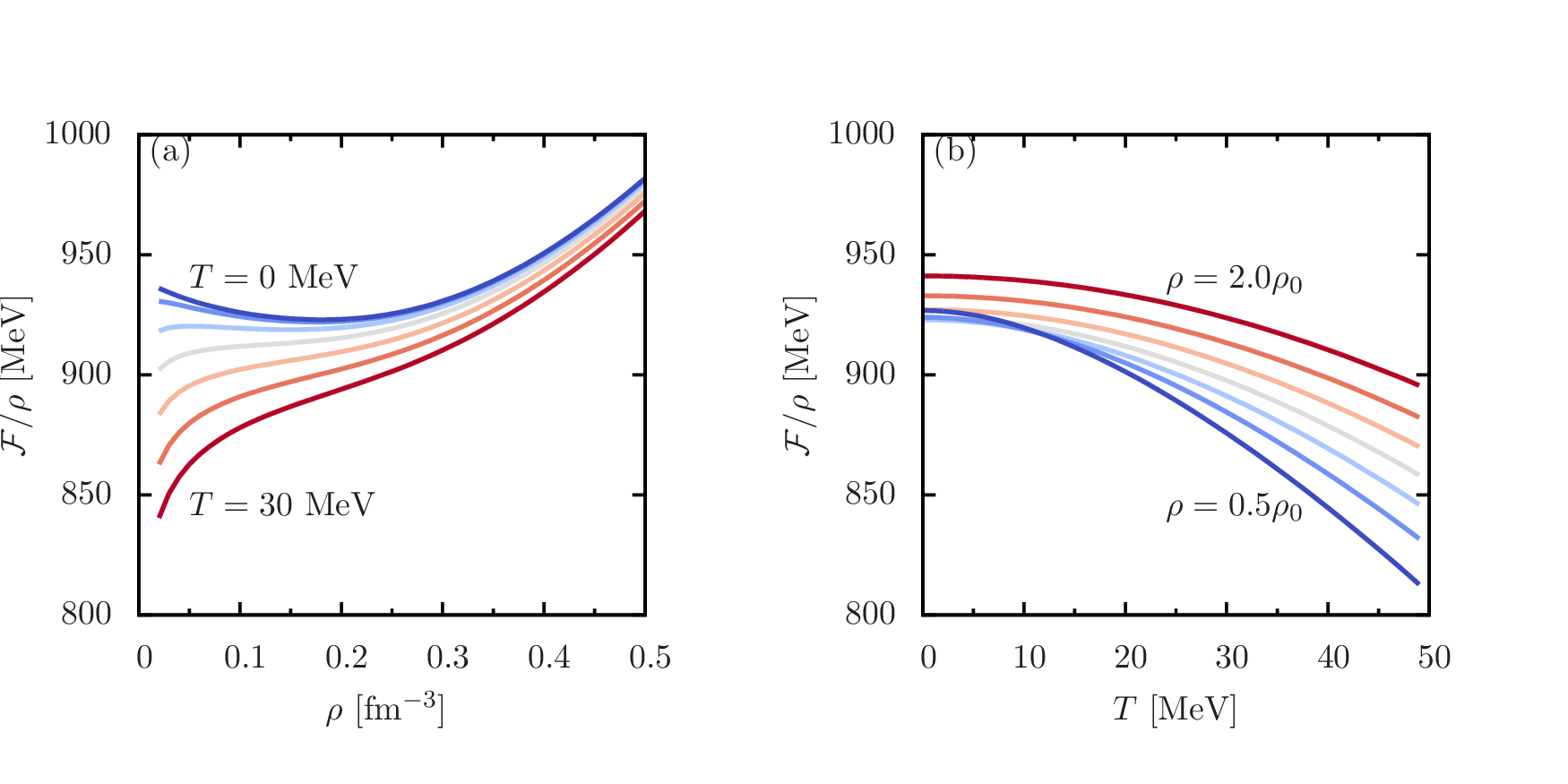}
	\caption{(Color online) The free energy per particle. Isotherms are chosen the
		same as Fig.~\ref{fig:Entropy-SymNM}.}
	\label{fig:FreeEnergy-SymNM}
\end{figure}

We conclude the set of results for symmetric nuclear matter with the discussion
of the free energy per particle,
$F/A=\mathcal{F}/\rho=(\mathcal{E}-T\mathcal{S})/\rho$.
Fig.~\ref{fig:FreeEnergy-SymNM} shows $\mathcal{F}/\rho$ as a function of
density and temperature.  At very low temperatures the free energy has a local
minimum, which is the equivalent to the saturation point of nuclear matter at
$T=0$.  At finite temperatures the equilibrium state of the system is given by
the minimization of the free energy  instead of the energy density. This minimum
describes a thermodynamically preferred state of the system.  The local minimum
of the free energy disappears above the so called flashing temperature,
$T_{\mathrm{F}}$.  At this point the pressure is still high enough to prevent
the system from decaying to the low density (gas) phase, leading to a liquid-gas
coexistence. Above a critical temperature, $T_C$, this coexistence does not hold
any longer and the system is found in the thermodynamically preferred state at
very low densities. For the DDRH model we find the following values for the
flashing and critical temperatures of symmetric nuclear matter:
$$ T^{\textrm{DDRH}}_F\approx12.2,\quad T^{\textrm{DDRH}}_C\approx14.6.$$

Although $E/A$ is more repulsive with rising temperature, the free energy per
particle is a decreasing function of $T$.
Obviously, this behavior can be ascribed to the $-T\mathcal{S}$ term and thus
is a pure entropic effect. The same applies to the  $\rho\rightarrow0$ limit,
where classical effects dominate the properties of the system (compare Fig.
\ref{fig:Entropy-SymNM}) \cite{Fiorilla:2012p4093}.

The analysis of symmetric nuclear matter shows that nucleon-nucleon interactions
affect the bulk properties of nuclear matter in two ways.  First, they obviously
give rise to the potential energy.  Second, the interactions invoke
self-consistent effective masses and chemical potentials. Especially the latter
quantities affect both the thermal and spectral distributions functions of the
nucleons. Therefore, the in-medium interactions have a remarkable  effect on the
kinetic part of the (free) energy, effectively increasing the kinetic energy.

In the considered density and temperature range, our results for the DDRH
model are in good agreement with experimental data as well as other microscopic
DBHF calculations \cite{Sammarruca:2010p4681}.  This shows, that the
approximations applied here to the density-dependent vertices are not only
suitable for the description of nuclear matter in the vicinity of the saturation
point at $T=0$, but lead also to reliable results in higher density as well as
temperature regions.

\section{Isospin Asymmetric Nuclear Matter} \label{sec:AsymNucMat}
\subsection{Thermal Properties of Asymmetric Nuclear Matter}
\begin{figure}[h]
	\centering
		\includegraphics{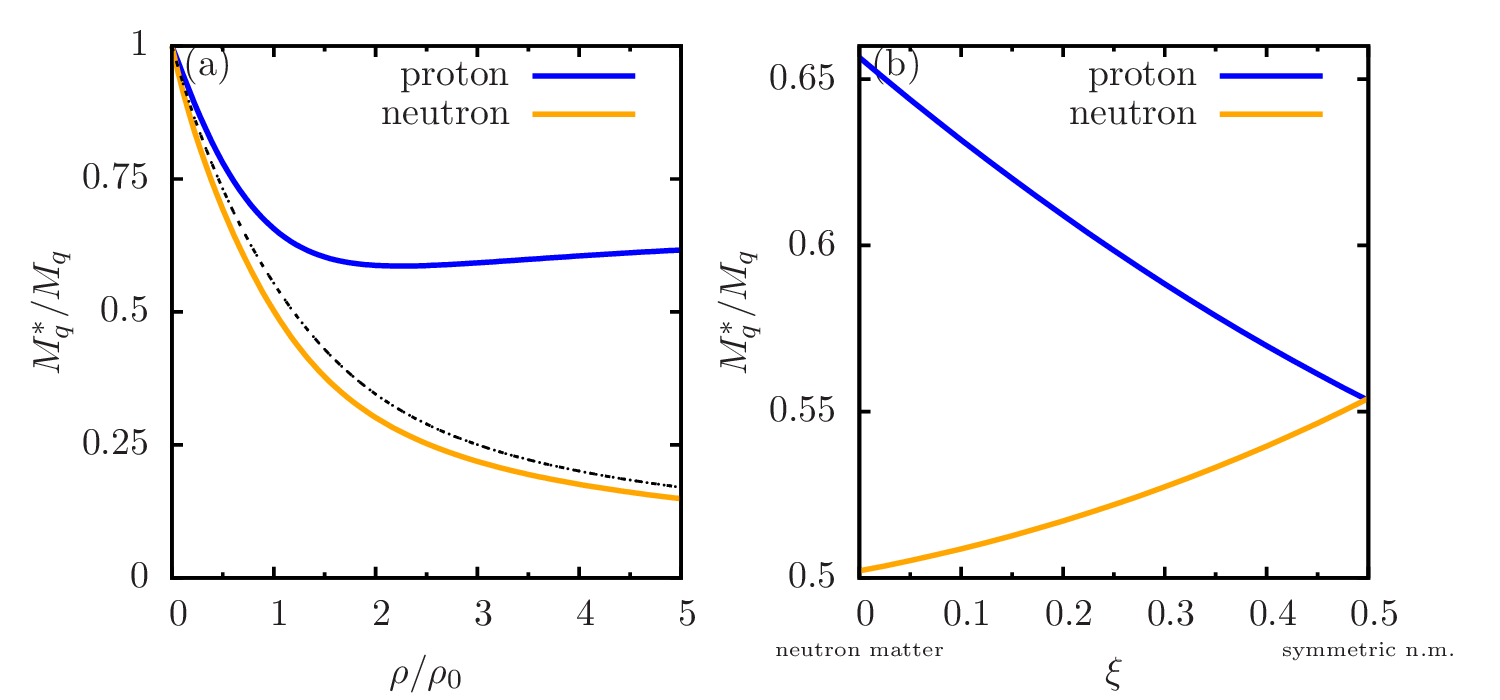}
		\caption{(Color online) Effective nucleon masses $M^*_q$ ($q=p,n$) for
			protons (upper blue) and neutrons (lower orange) at $T=0$ as functions of
			the total baryon density for pure neutron matter (left) and as functions
			of the asymmetry parameter $\xi=\frac{Z}{A}$ at saturation density
		$\rho=\rho_0$ (right). The thin dashed curve in the left panel represents
	the result in case of symmetric nuclear matter.}
	\label{fig:Mstar-Xsi}
\end{figure}

In the previous section we studied the results for nuclear matter with equal
content of protons and neutrons. In this section we study isospin effects in
warm nuclear matter at moderate temperature well below the critical temperature
$T_C$ of the QCD phase transition. For this purpose, we introduce the proton
fraction $\xi\equiv\rho_p/\rho$ as an indicator for charge asymmetry and the
isospin content.  In the DDRH model, the inclusion of the scalar isovector
$\delta$ meson field leads to a separation of the proton and neutron effective
masses in asymmetric nuclear matter. In Fig.~\ref{fig:Mstar-Xsi} we plot the
ratio between the effective and the corresponding bare nucleon mass.  The black
dashed curve on the left panel shows the result for symmetric nuclear matter
($\xi=0.5$). In this case $M^*_p=M^*_n$ decreases continuously with $\rho$.  As
the asymmetry increases (i.e.\ smaller values of $\xi$), protons gain a larger
effective mass, while the effective mass of neutrons decreases.  It is
particularly interesting to note, that the isospin effect on $M^*_p$ is much
stronger than on $M^*_n$.  In the limit of $\xi=0$ (neutron matter) $M^*_n$
deviates only slightly from its symmetric nuclear matter value, whereas $M^*_p$
shows a considerable increase and even exhibits a local minimum at $\rho\approx
2.3\rho_0$.  This behavior is a direct consequence of the self-consistent
solution of the effective masses and scalar densities.

\begin{figure}[h]
	\centering
		\includegraphics{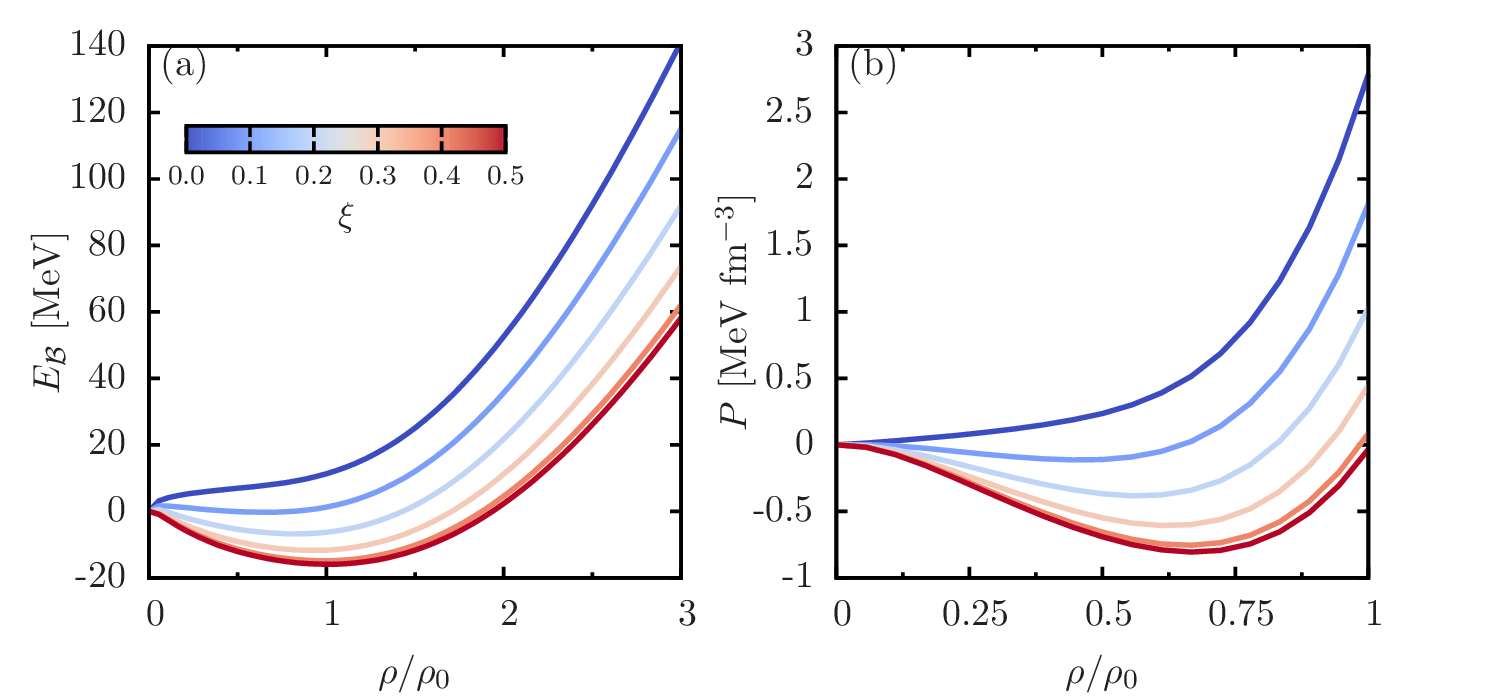}
		\caption{(Color online) The binding energy (left) and the pressure (right)
		at $T=0$ for values of $\xi$ from $0$ to $.5$ in steps of 0.1. Lower curves
		belong to lower values of $\xi$.}
	\label{fig:EoS-Asym}
\end{figure}

Fig.~\ref{fig:EoS-Asym} shows the energy density and pressure at $T=0$ for
values of $\xi$ from pure neutron matter ($\xi=0$) to symmetric matter
($\xi=0.5$. As seen, the local minimum of the binding energy per particle shifts
to the left, i.e. to lower densities, while the binding of the system reduces
with increasing neutron excess. For $\xi<.05$ neutron-rich matter is completely
unbound. At this point the potential energy is not strong enough to compensate
the kinetic energy of the system. Hence, both, energy density and pressure
increase with decreasing proton fraction. In neutron-rich matter, the local
minimum of the pressure disappears as well, as the right panel of 
Fig.~\ref{fig:EoS-Asym} shows. Thus, the thermodynamically preferred state of
neutron matter is a dilute Fermi gas in the entire density range.

\begin{figure}[h]
	\centering
		\includegraphics{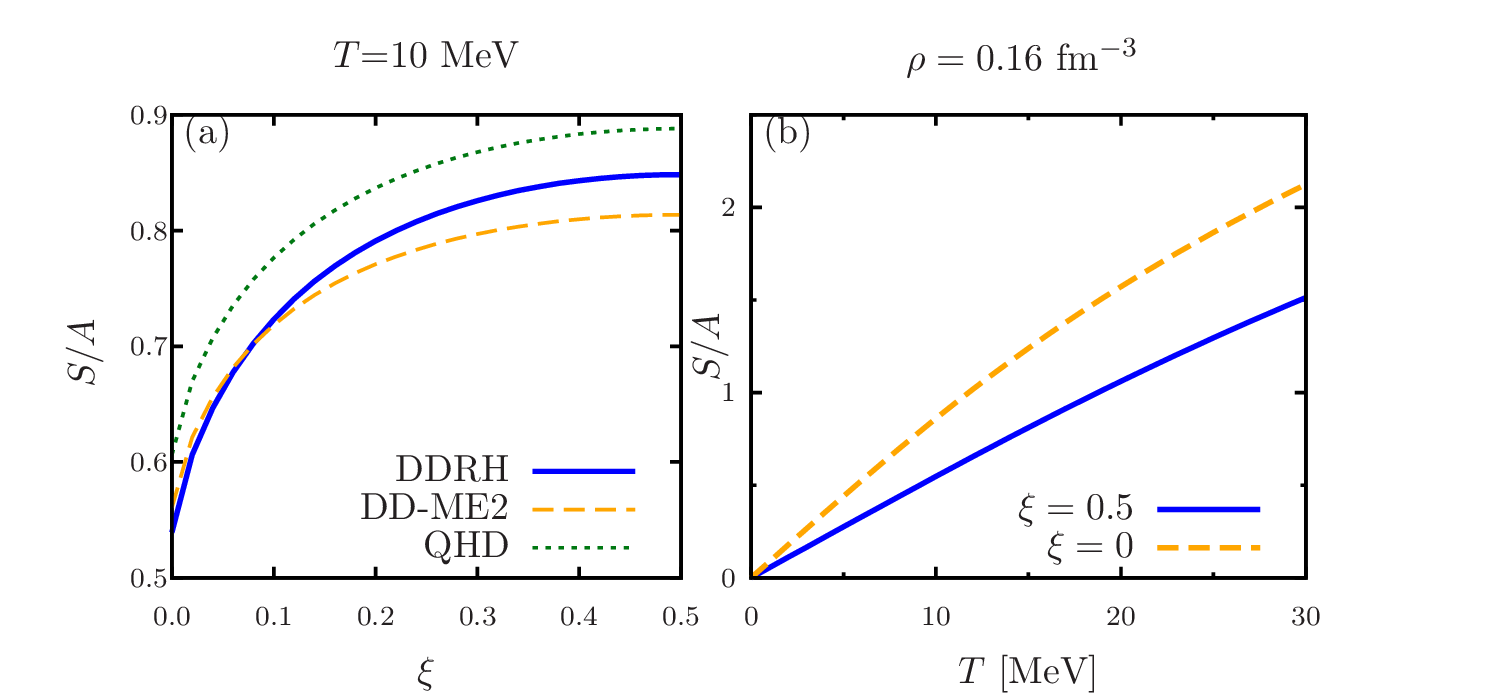}
		\caption{(Color online) Left: The entropy per particle as a function of the
			proton fraction $\xi$ at $T$=10 and $\rho$=0.16 fm$^{-3}$. The results for
			the DDRH, DD-ME2 \cite{Lala:2005} and QHD parametrization are shown. Right:
			Comparison of the temperature dependence of the entropy per particle
			between symmetric (upper curve) and pure neutron matter (lower curve) in
			the DDRH model at fixed density $\rho=0.16$ fm$^{-3}$. }
	\label{fig:EntropyAsymNM}
\end{figure}

The entropy $S/A$ decreases for smaller values of $\xi$ as the degrees of
freedom are reduced in systems with smaller proton fractions.  In our
calculations, we find a very similar asymmetry dependence of $S/A$ in the whole
temperature range.  As an example, we show $S/A$ as a function of $\xi$ within
several models at a fixed temperature of $T=10$ MeV. The results show that the
calculation with the linear QHD parametrization produces 
somewhat larger values of the entropy.  The blue solid curve of the DDRH
calculation lies above the results with the phenomenological DD-ME2 
parametrization \cite{Lala:2005}.  This difference is less pronounced at 
smaller $\xi$, however, and at neutron matter the entropy
in the DDRH model is even slightly smaller than the one of the DD-ME2 calculation.
In the right panel of Fig.~\ref{fig:EntropyAsymNM} we compare the entropy per
particle as a function of the temperature between symmetric nuclear matter and
neutron matter within the DDRH model. For neutron matter $S/A$ shows a linear
temperature dependence throughout almost the whole temperature range.

\begin{figure}[h]
	\centering
		\includegraphics{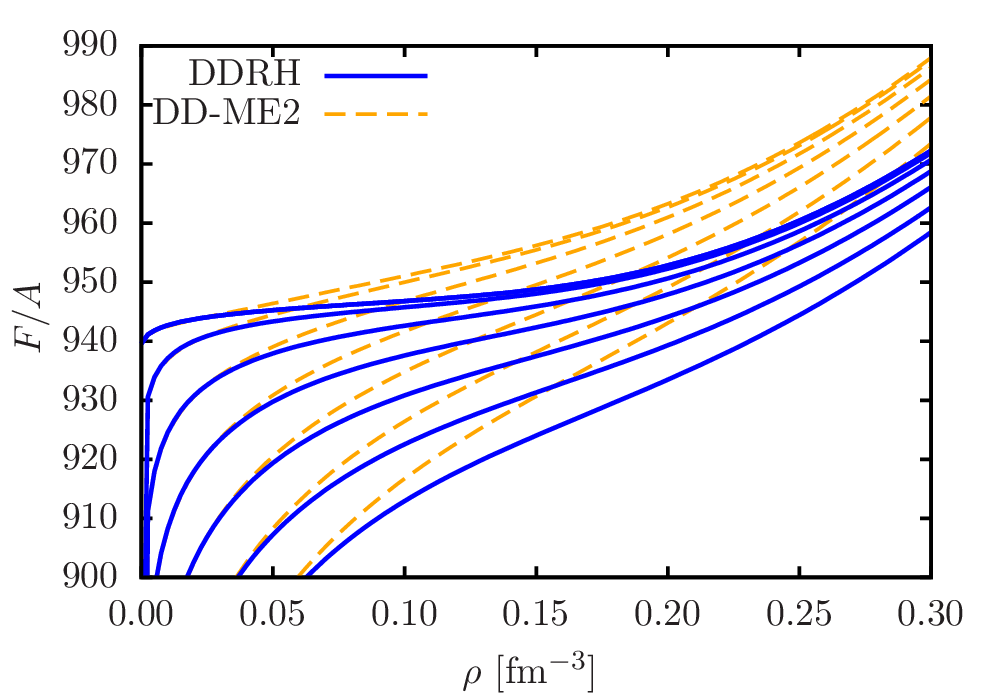}
		\caption{(Color online) Free energy per particle in neutron matter for temperatures from 0
			to 30 MeV in equidistant steps of 5 MeV (lower curves correspond to higher
			temperatures).
      The results of the DDRH (solid lines) and the DD-ME2 (dashed lines) model are compared.}
	\label{fig:FreeEnergyAsymNM}
\end{figure}

In Fig.~\ref{fig:FreeEnergyAsymNM} the free energy per particle of neutron
matter is compared between the DDRH and the phenomenological DD-ME2 model.
The results show that the DDRH parametrization leads to a less repulsive free
energy which is a consequence of the additional attraction in the
scalar-isovector channel.  At low densities the curves of the two models
coincide with each other, since the differences between the interactions become
less important in this density region.  As in the case of symmetric nuclear
matter, $F/A$ decreases with temperature due to the higher values of the
entropy.

To investigate the isospin dependent part of the equation of state, consider the
expansion of the free energy in powers of the isospin parameter
$\alpha\equiv(\rho_n-\rho_p)/\rho$ around symmetric nuclear matter ($\alpha$=0):
\begin{align}
  F(\alpha,\rho,T)=F(0,\rho,T) + F_\mathrm{sym}(0,\rho,T)\alpha^2+\mathcal{O}(\alpha^4).
\label{eq:SymEnDef}
\end{align}
Because of isospin symmetry, odd powers of $\alpha$ vanish in the above
expansion.
Eq.~\ref{eq:SymEnDef} defines the {free symmetry energy}, $F_\mathrm{sym}$,
which is related to the cost of converting protons into neutrons in the nuclear
medium.
Accordingly, the {symmetry energy}, $E_\mathrm{sym}$, and the {symmetry
entropy}, $S_\mathrm{sym}$, can be defined in the same way.
The relation to the proton fraction parameter $\xi$ is then given by
\begin{align}
	A_\mathrm{sym}=\frac{1}{8}\frac{1}{\rho}\frac{\partial^2 \mathcal{A}}{\partial\xi^2}\Bigg|_{\xi=0.5},\quad\mathcal{A}\in\left\{ \mathcal{F},~\mathcal{E},\mathcal{S} \right\}
\end{align}
As usual, the above symmetry functionals follow the general thermodynamic
relation $F_\mathrm{sym}=E_\mathrm{sym}-TS_\mathrm{sym}$.

The symmetry energy plays a fundamental role in the description of many
important phenomena in nuclear physics, such as the structure of exotic nuclei,
heavy ion collisions and neutron stars.
In the past decade the density  dependence of the symmetry energy has been
extensively studied both experimentally and theoretically \cite{Li:2008p5018}.
In addition, the knowledge of the temperature dependence of the (free) symmetry
energy has become more and more important in connection with the analysis of
multi-fragmentation data of hot nuclear matter. It is also an important
ingredient in astrophysical calculations such as the evolution of proto-neutron
stars or the core collapse of a massive star and the associated explosive
nucleo-synthesis \cite{Janka:2007p5024, Roberts:2012p5025}.
In the last years some progress has been achieved in heavy ion collision
experiments to extract the temperature dependence of $F_\mathrm{sym}$ at low
densities \cite{Wada:2012p5023}.

\begin{figure}[h]
	\centering
	\includegraphics{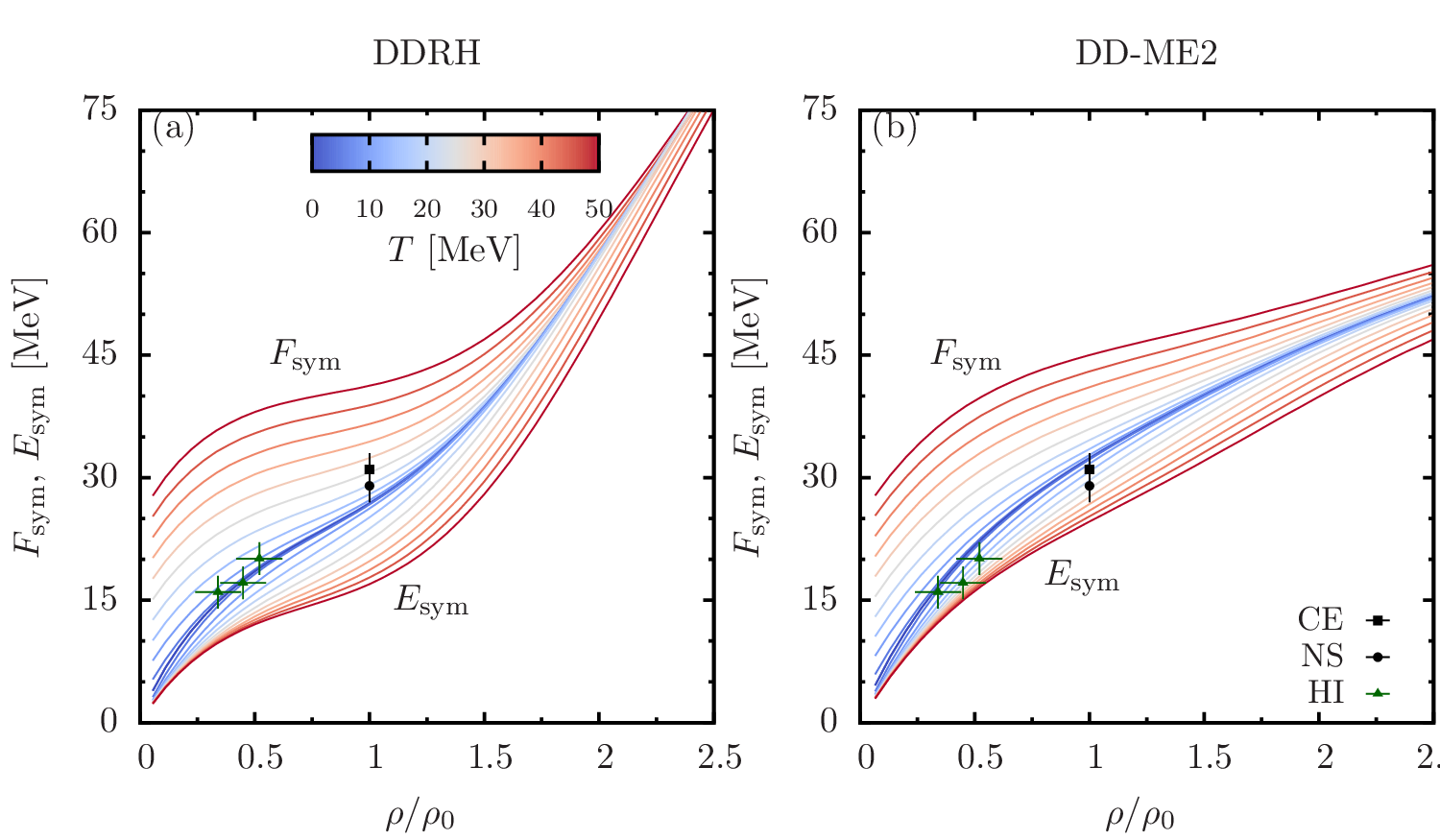}
		\caption{(Color online) Symmetry and free symmetry energy as functions of
			the density for fixed values of $T$ from 0 to 50 MeV.
			At $T$=0 (blue curve) $F_\mathrm{sym}=E_\mathrm{sym}$. Curves above
			(below) the zero temperature line correspond to $F_\mathrm{sym}$
			($E_\mathrm{sym}$). The right and the left panel show the results for the
		DDRH and DD-ME2 model, respectively. The points indicate experimental results
	for the symmetry energy at $T$=0. The data points are taken from \cite{Khoa:2005} (CE)
	, \cite{Furn:2002} (NS) and \cite{Shetty:2007} (HI)}
	\label{fig:SymmetryEnergy}
\end{figure}

In Fig.~\ref{fig:SymmetryEnergy} we show our prediction for $F_\mathrm{sym}$ and
$E_\mathrm{sym}$ (see also Table \ref{tab:SymmetryEnergy}). Again, results for 
DDRH and DD-ME2 parameter sets are compared
together with the experimental data points from charge exchange reactions 
\cite{Khoa:2005}, neutron skin analysis \cite{Furn:2002} and heavy ion 
collisions  \cite{Shetty:2007}. 
The symmetry energy of the DDRH model shows a stiff density dependence, while it
is soft for the phenomenological DD-ME2 model parameter set. At $T$=0 and at low
densities both models describe the experimental data quite well.
At saturation density the DDRH parametrization slightly underestimates the
experimental value for $E_\mathrm{sym}$.
However, since -- different to the DD-ME2 model -- the DDRH parameter set has not
been fitted to experimental data this result is already quite noteworthy. As for
the temperature dependence, one can see that $F_\mathrm{sym}$ and
$E_\mathrm{sym}$ show an opposite behavior. While $F_\mathrm{sym}$ increases
with $T$, $E_\mathrm{sym}$ becomes smaller with rising temperature.
The decrease of $E_\mathrm{sym}$ at higher temperatures can be understood from
the fact that the Fermi surface is more diffuse and therefore the Pauli blocking
becomes less important at increasingly higher temperatures.
The increase of the free symmetry energy, however, is related to the negative
value of $S_\mathrm{sym}$, since the total entropy per particle decreases with
increasing asymmetry (Fig.~\ref{fig:EntropyAsymNM}). Consequently,
$F_\mathrm{sym}$ is expected to be larger than $E_\mathrm{sym}$ at fixed density
and temperature values. Additionally, the entropic effect of the free symmetry
energy is stronger than the decrease of $E_\mathrm{sym}$ which all in all causes
$F_\mathrm{sym}$ to increase with $T$.
\begin{table}
	\caption{\label{tab:SymmetryEnergy} $E_\mathrm{sym}$ and $F_\mathrm{sym}$ taken at saturation density for different values of temperature $T$ in the DDRH and DD-ME2 model. }
	\begin{ruledtabular}
		\begin{tabular}{lllll}
			T [MeV]& $E_\mathrm{sym}^\mathrm{DDRH}$ [MeV] & $E_\mathrm{sym}^\mathrm{DD-ME2}$ [MeV]& $F_\mathrm{sym}^\mathrm{DDRH}$ [MeV]& $F_\mathrm{sym}^\mathrm{DD-ME2}$[MeV]\\
			\hline
			10  & 26 & 32 & 27 & 33\\
			20 & 24 & 30 & 29 & 35 \\
			30 & 21 & 28 & 32 & 37 \\
			40 & 19 & 26 & 36 & 41\\
			50 & 17 & 25 & 41 & 45\\

	 \end{tabular}
 \end{ruledtabular}
\end{table}

\subsection{Phase Transitions in Asymmetric Nuclear Matter}
As a starting point, let us first recall some basic concepts of phase
transitions in nuclear matter.
The equation of state of nuclear matter shows a typical van-der-Waals gas
behavior at temperatures $0<T<20$ MeV \cite{Jaqaman:1983p4276}. Below a critical
temperature  $T_C$ one finds a region where $\frac{\partial P}{\partial
\rho}\Big|_T<0$.
This is the region of mechanical instability where the preferred state of the
system is given by two coexisting phases.
The thermodynamical equilibrium of the two phases can be obtained with the help
of the Gibbs-Duhem relation,
\begin{align}
	\td P -\mathcal{S}\td T - \sum_c \rho_c \td \mu_c=0,
	\label{eq:Gibbs-Duhem}
\end{align}
where the sum includes all conserved charges. Eq.~\eqref{eq:Gibbs-Duhem} applies
to both phases separately. In case of one conserved charge this leads to the
Gibbs conditions
\begin{align}
	\mu^{I}(\rho^{I},T)=\mu^{II}(\rho^{II},T),\\
	P^{I}(\rho^{I},T)=P^{II}(\rho^{II},T),
	\label{eq:Gibbs}
\end{align}
where the labels $I$ and $II$ refer to the two phase-states, with the convention
$\rho^{I}<\rho^{II}$.
Recalling that the preferred state of a system is the one with the lowest value
of the free energy, we can write down the global stability condition for the
mixed phase \cite{Muel:1995}:
\begin{align}
	\mathcal{F}(\rho, T) < \lambda \mathcal{F}(\rho^{I},T)+(1-\lambda)\mathcal{F}(\rho^{II},T),
	\label{eq:Fconv}
\end{align}
with
\begin{align}
	\nonumber
	\rho=\lambda\rho^{I}+(1-\lambda)\rho^{II}\quad \lambda\in[0,1],
\end{align}
where the parameter $\lambda$ determines the volume fraction occupied by each
phase.  Eq.~\eqref{eq:Fconv} implies that  for a stable system $\mathcal{F}$
should be a convex function of the density,
\[\frac{\partial^2\mathcal{F}}{\partial \rho^2}\geq 0.\]
Note, that the above expression is equivalent to $\frac{\partial \mu}{\partial
\rho}\geq 0$, as can be easily deduced from \(\td\mathcal{F}=-\mathcal{S}\td T
+\sum_c \mu_c\mathcal\td \rho_c\).
With this, the phase transition region is characterized by the following states,
\begin{itemize}
	\item{\em spinodal curve}: describes the onset of the instability region, $\frac{\partial^2\mathcal{F}}{\partial \rho^2}=0$
	\item{\em metastable region}: the stability conditions are fulfilled locally, but Eq.~\eqref{eq:Fconv} is violated
	\item{\em binodal curve}: describes the onset of the metastable states.
\end{itemize}

\begin{figure}[h]
		\centering
		\includegraphics{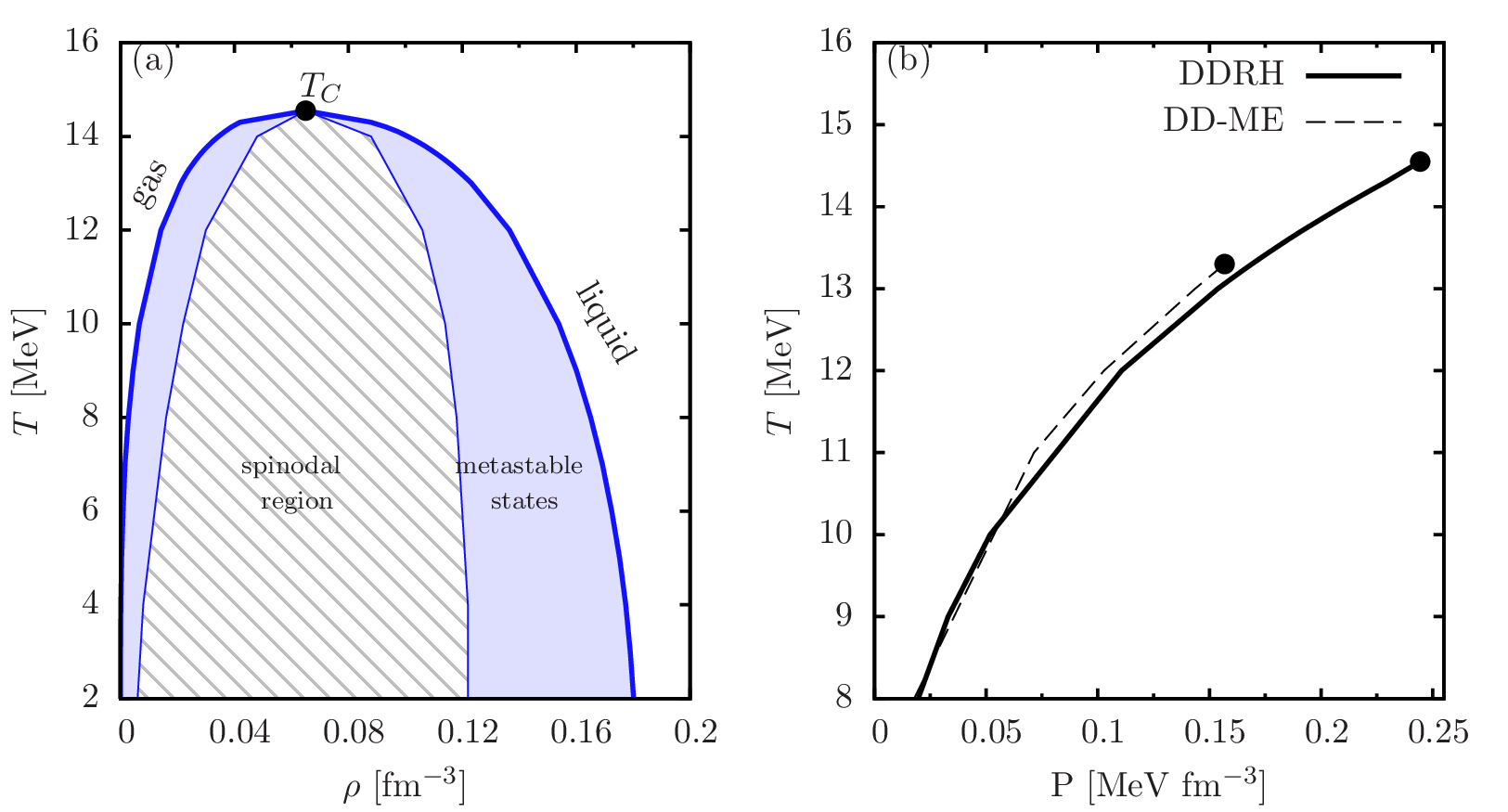}
		\caption{(Color online) The phase transition diagrams of symmetric nuclear matter. Left:
		The $T$-$\rho$ phase diagram in the DDRH model. The binodal and spinodal
	curves are indicated by thick and thin lines, respectively. Right: The
$T$-$P$ phase diagram for the DDRH and DD-ME2 model calculations.}
		\label{fig:Binodal-DDRH}
\end{figure}

Hence, the two-phase coexistence area is enclosed by the binodal curve. Within
this area the pressure and the chemical potentials are kept constant. The
projection of the coexistence area onto the $T$-$P$ or $T$-$\mu$ plane results
in the first-order phase transition lines.
Together with the $T$-$\rho$ phase diagram they provide a full description of the phase transition.
In Fig.~\ref{fig:Binodal-DDRH} we present the DDRH results of the $T$-$\rho$
(left)  and the $T$-$P$ (right) phase diagrams.  The hatched area indicates the
region of mechanical instability.  We find a critical temperature of $T_C\approx
14.55$ MeV.  The calculation with the DD-ME2 parameter set provides similar
results, although the values of the critical points are a bit smaller than in
the DDRH case.
\\
The above conditions can be generalized to describe phase transitions of
multi-component systems with $N$ conserved charges, e.g. hyper nuclear matter.
The stability criterion \eqref{eq:Fconv} then holds for each conserved charge
density $\rho_i$. The resulting set of $N$ inequalities implies
\cite{Muel:1995}
\begin{align}
	\frac{\partial\mu_i}{\partial\rho_j}=\frac{\partial\mu_j}{\partial\rho_i}>0
\end{align}
In the coexistence region the Gibbs conditions of a multi-component system have
to be fulfilled for each particle type,
\begin{align}
	\mu_i(\rho_i^{I})=\mu_i(\rho_i^{{II}}),\quad P(\rho_i^{I})=P(\rho_i^{{II}}).
\end{align}
For isospin asymmetric nuclear matter the conserved charges of the system are
given by the baryon number $B$ and the third component of the total isospin
$I_3$.  Any state of the nuclear matter system can thus be characterized by the
baryon density $\rho_B$=$\rho_p+\rho_n$ and the isovector density
$\rho_3=\rho_p-\rho_n$.  The corresponding baryon and isospin chemical
potentials are related to the nucleon and proton chemical potentials through
\[ \mu_B=\mu_p+\mu_n,\quad \mu_3=\mu_p-\mu_n. \]
It is feasible to express the stability conditions in terms of the proton fraction $\xi$,
\begin{align}
	\rho\left(\frac{\partial P}{\partial\rho}\right)_{T,\xi}>0,\quad\left( \frac{\partial\mu_p}{\partial\xi} \right)_{T,P}>0,\quad\left( \frac{\partial\mu_n}{\partial \xi}\right)_{T,P}<0.
\end{align}
The last two expressions are referred to as \emph{chemical stability}
conditions.  They take into account the fact that there is energy needed to
change the concentration of protons in the medium at a fixed temperature and
pressure.
\begin{figure}[ht]
		\centering
		\includegraphics{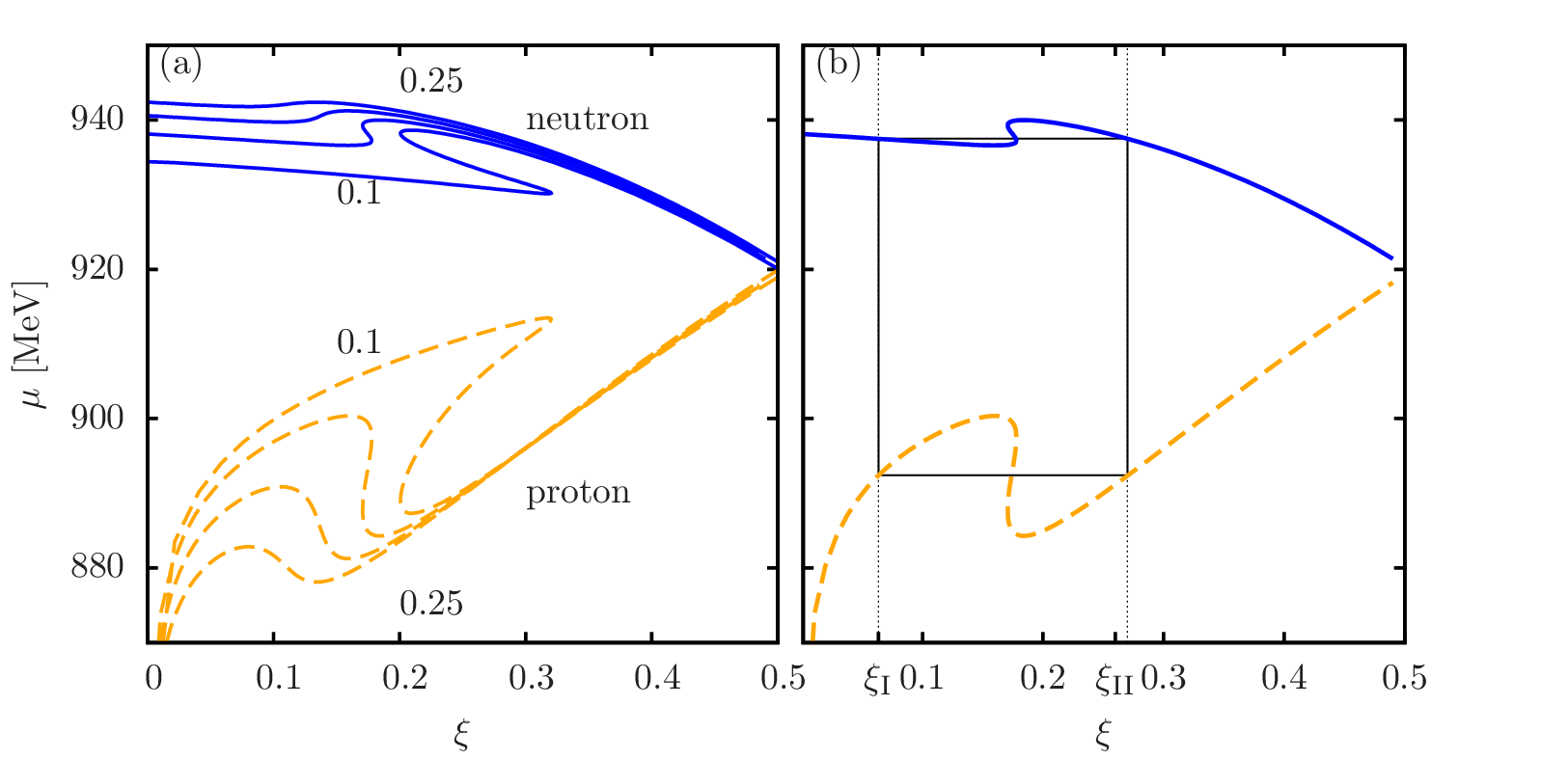}
		\caption{(Color online) The chemical potential isobars of nuclear matter in
			the DDRH model as a function of $\xi$ at $T=10$ MeV. The left panel shows
			the neutron (solid blue) and proton (dashed orange) isobars for values of
			$P$ from 0.1 to 0.25 MeV fm$^{-3}$. The right panel indicates the
			geometrical construction in case of $P$=0.15 MeV fm$^{-3}$ as described in
		the text.}
\label{fig:Isobars}
\end{figure}
The chemical instability region is found from the analysis of the neutron and
proton chemical potential isobars as functions of $\xi$.  In
Fig.~\ref{fig:Isobars} a set of isobars in the range of $0.1$ MeV~fm$^{-3}
<P<0.25$~MeV~fm$^{-3}$ is shown. One can see an area in the  $P$-$\xi$ space,
where the chemical stability conditions are violated.
In this section the system would break apart into two phases with different
concentrations of its constituents. With increasing pressure the instability
region becomes smaller until it disappears completely at the critical pressure
$P_C$. At this point an inflection point appears in the chemical potential
isobars,
\[\left( \frac{\partial\mu}{\partial\xi} \right)_{T,P=P_C}=\left(\frac{\partial^2\mu}{\partial\xi^2} \right)_{T,P=P_C}=0\]
In models with constant nucleon-meson couplings the position of the inflection
points of protons and neutrons coincides in the $P$-$\xi$ plane.  This
assumption is, however, no longer true, once the couplings become density
dependent.  In this case, the chemical potential of one type of nucleons can
pass an inflection point while the other one still has an unstable region.  This
circumstance was first found by Qian \cite{Qian:2000p5052} where it was
shown that the asynchronous behavior of the nucleon specie varies according to
the density dependence of the isovector-vector coupling, $\Gamma_\rho$.  In
agreement with that, we find a similar situation for the DDRH model.

In phase equilibrium, the pressure and the chemical potentials of the two phases
are equal, as required by the Gibbs conditions.  The two solutions to this
requirement can be found by means of a geometrical construction, as shown in the
right panel of Fig.~\ref{fig:Isobars} \cite{Lattimer:1978p5047}.  The points
$\xi_\textsc{i}$ and $\xi_\textsc{ii}$ indicate the onset of the coexistence
region.  Thus, owing to the additional degree of freedom, the binodal curve
becomes a surface in the $T$-$P$-$\xi$ space.  To better visualize this surface,
it is usually displayed in slices at constant $T$, which is referred to as the
\emph{binodal section}.  The shape of the binodal section depends highly on the
model of the nucleon-nucleon interaction.  

\begin{figure}[h]
		\centering
		\includegraphics{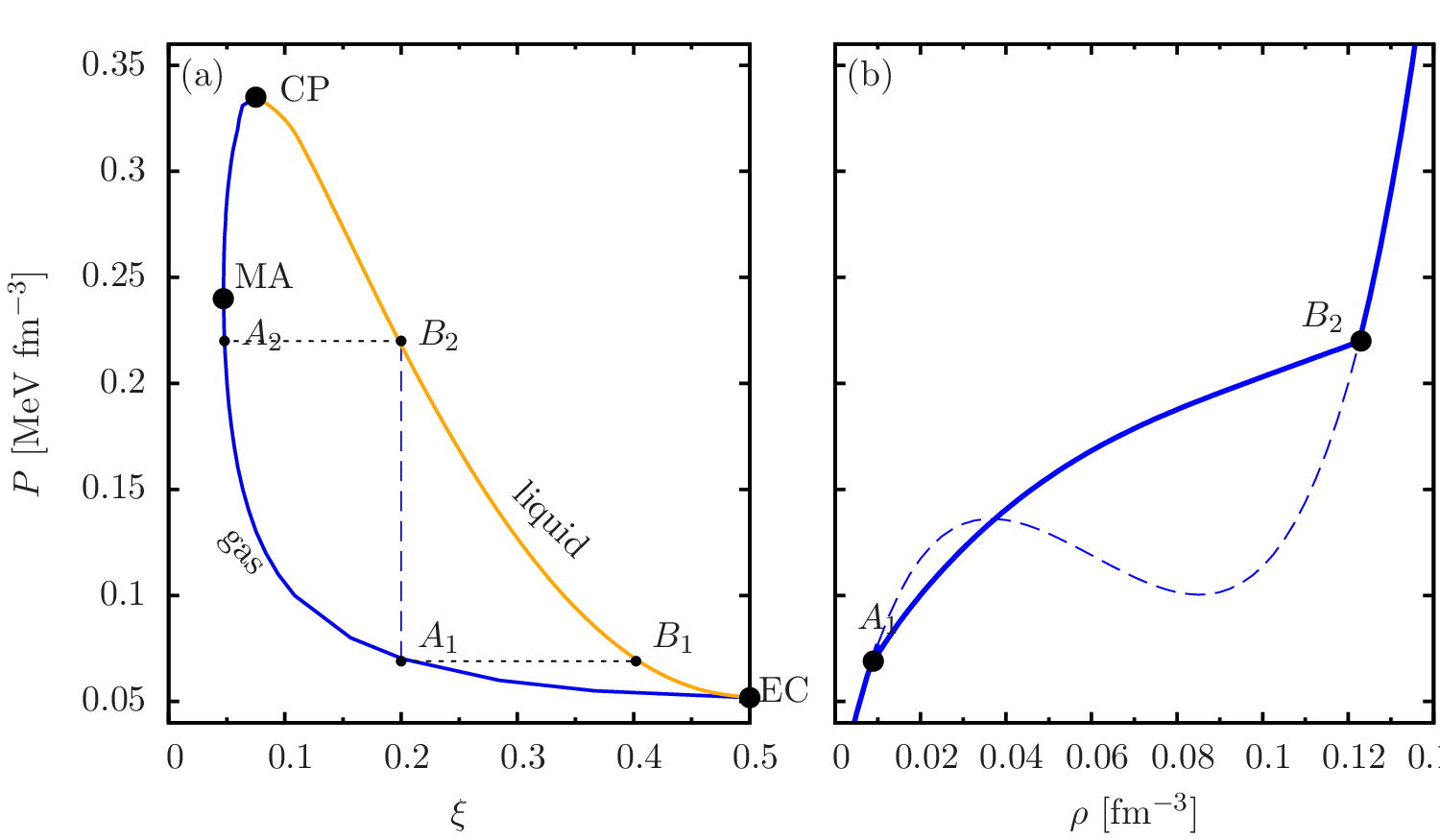}
		\caption{(Color online) The binodal section at $T$=10 MeV in the DDRH model. The right
		panel shows the projection of the points $A_1$ and $A_2$ on the pressure at
	fixed proton concentration $\xi=0.2$.}
\label{fig:Binodal-Section}
\end{figure}

To illustrate the nature of the liquid-gas transition in asymmetric nuclear
matter, we show  the result of the binodal section of the DDRH model at $T$=10
MeV in Fig.~\ref{fig:Binodal-Section}.  Following the notation of
\cite{Muel:1995}, we indicate some characteristic points of the curve. By
\textit{MA} the point at maximum asymmetry is indicated. For
$\xi<\xi_\textmd{MA}$ the system stays in the gas phase and a phase transition
does not take place. The point of equal concentration, \textit{EC}, lies at
$\xi=0.5$, reflecting the one-component character of symmetric nuclear matter.
The critical point \textit{CP}($\xi_C$, $P_C$) indicates the edge of the
instability area. For $P>P_C$ the chemical potentials of protons and neutrons
are monotonically with $\xi$. 

In the phase coexistence region the system favors a configuration with different
proton concentrations of the two phases. This does not violate the
isospin conservation law, since only the sum of the isospin of the two phases
needs to be conserved. One finds, that the phase with lower $\xi$ (higher
asymmetry) takes the lower density.  The points \textit{CP} and \textit{EC}
divide the binodal section into two branches, where the left(right) branch
corresponds to the phase with lower(higher) density and proton concentration.
Thus, the left branch is associated with the gas phase and the right one with
the liquid phase. The binodal section shrinks with temperature, i.e.,
$P_{\mathrm{CP}}$ becomes smaller and \textit{MA} shifts to higher values of
$\xi$. At $T$=$T_C$ the points  \textit{CP}, \textit{MA} and \textit{EC}
coincide at $\xi=\frac{1}{2}$ and the binodal surface finally reduces to a
single point.

In contrast to symmetric nuclear matter, the pressure does not stay constant
during a phase transition. This arises from the fact that the two phases follow
the two different branches along the binodal curve.  As an example, consider a
system in a configuration below the binodal section with $\xi$=0.2.  As the
system is isothermally compressed, it will reach the two-phase instability
region at the point $A_1$.  In this stage a second phase emerges at the point
$B_2$ with $\xi\approx$0.4 (liquid phase).  During the phase transition the
total proton fraction $\xi$ is held fixed, as indicated by the vertical line in
the left panel of Fig.~\ref{fig:Binodal-Section}.  The two coexisting phases
evolve along the two different branches of the binodal section.  The gas phase
follows the left branch from $A_1$ to $A_2$ and the liquid phase follows the
right branch from $B_1$  to $B_2$. Finally, the system leaves the instability
area at the point $B_2$.
The solution of the equations
\[
	\rho=\lambda \rho^\textsc{i}+(1-\lambda)\rho^\textsc{ii}	
\]
\[
	\rho_3=(2\xi-1)\rho=\lambda \rho_3^\textsc{i}+(1-\lambda)\rho_3^\textsc{ii}	
\]
provides the fraction $\lambda$ of the volume  which is occupied by the gas phase.
In the above example, $\lambda$=1 at $A_1$ and vanishes at $A_2$, implying that
the system evolves from a gas to a liquid phase.
The solution of the above equations can be used to calculate the behavior of the
pressure during the phase transition.
This is illustrated in the right panel of Fig.~\ref{fig:Binodal-Section}.

The behavior of the system during an isothermal compression can be very
different depending on the asymmetry parameter. In general, one can    
distinguish between the following two condensation types

\begin{itemize}
	\item $\xi>\xi_{\mathrm{s}}$: stable condensation \\
		starting in the gas phase, the system undergoes a phase
		transition and ends in the stable liquid phase.
\item $\xi<\xi_{s}$: retrograde condensation \\
	The system starts and terminates its evolution through the two-phase
	coexistence region in the gas phase.
	The liquid phase which emerges during the transition disappears again as the
	upper boundary of the binodal section is reached.
	This behavior does not occur in a one-component system.
\end{itemize}

\begin{figure}[h] \centering \includegraphics{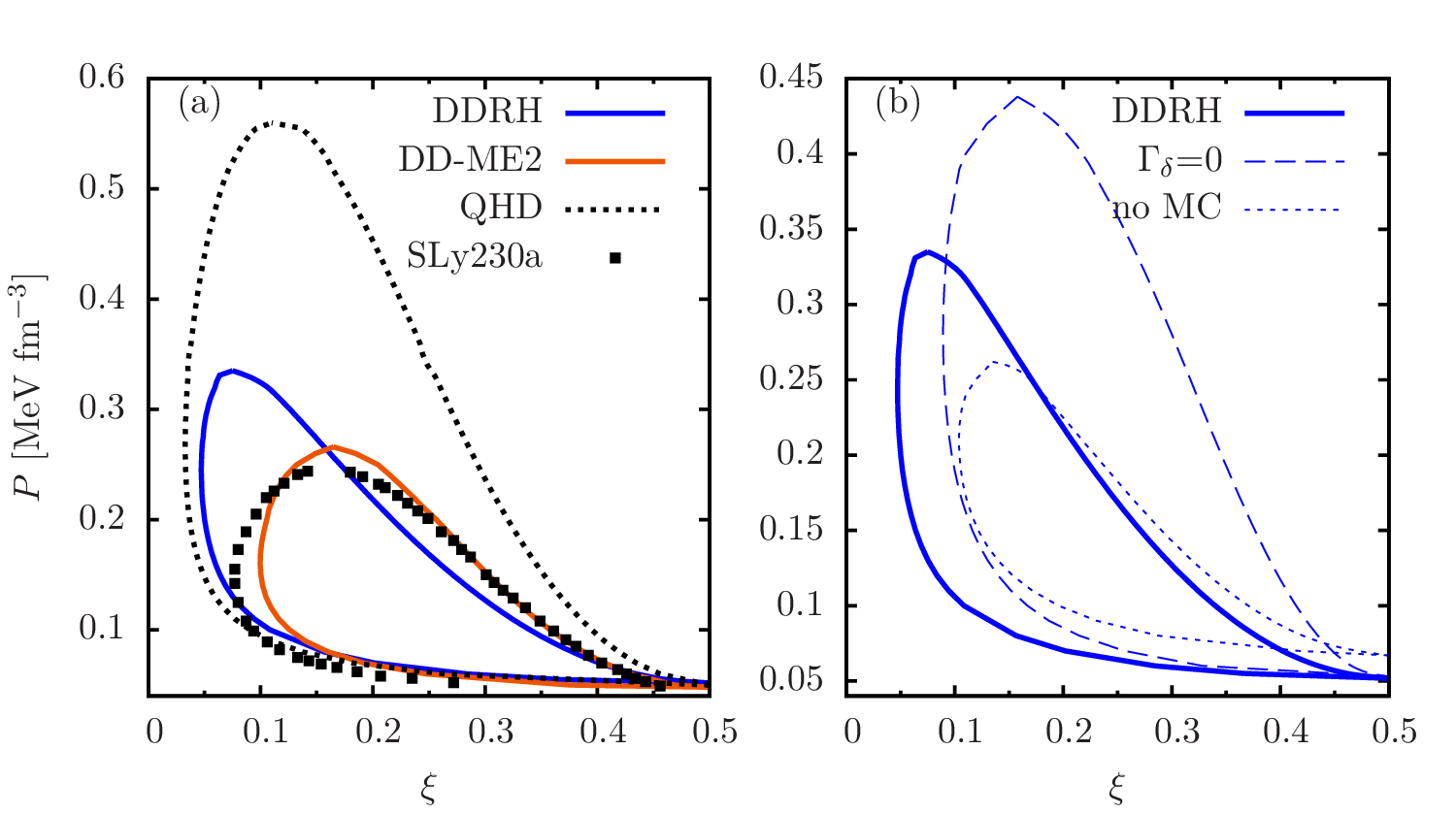}
	\caption{(Color online) (a) Comparison of the binodal surface at $T$=10 MeV
		between the DDRH, DD-ME2, QHD and SLy230a \cite{Dutra:2008} model. (b) Results of calculations without
		$\delta$ meson interaction (dashed) and without momentum correction (short
	dashed) are compared to the full DDRH calculation.}
	\label{fig:Binodal-AsymNM-Models} \end{figure}

It is interesting to examine, how the shape of the binodal section depends on
the nucleon-nucleon interaction.  Especially the position of the characteristic
points is very sensitive to the isospin part of the interaction. In Fig.
~\ref{fig:Binodal-AsymNM-Models}(a) the binodal section at $T$=10 MeV of the
DDRH model is compared to the results from other models. A summary of the
corresponding critical values can be found in Tab.~\ref{tab:BinodalPoints}.

\begin{table}
	\caption{\label{tab:BinodalPoints} 
	Comparison of the critical point parameters for several temperatures.   The
	results for the NL3 and SLy230a models are taken from \cite{Dutra:2008}
}
	\begin{ruledtabular}
		\begin{tabular}{llll}
			\quad & T [MeV]	& $\xi_{c}$ & $\rho_c$ [fm$^{-3}$] \\
			\hline
			DDRH & 0 & 0.0135 & 0.059 \\
			DD-ME2 & 0 & 0.0025 & 0.0327 \\
			NL3 & 0 &  0.0567 & 0.0766 \\
			SLy230a & 0 & 0.0149 & 0.0843 \\
			\\
			DDRH & 10 & 0.077  & 0.079   \\
			DD-ME2 & 10 & 0.167 & 0.065  \\
			NL3 & 10 & 0.1785 &  0.0573   \\
			SLy230a & 10 & 0.110 & 0.0608   \\
			\\
			DDRH & 14.55  & 0.5 & 0.0653 \\
			DD-ME2 & 13.31 & 0.5 & 0.045 \\
			NL3 & 14.55 & 0.5 & 0.0463\\
			SLy230a & 16.52 & 0.5 & 0.0535 \\
	 \end{tabular}
 \end{ruledtabular}
\end{table}

As in the case of symmetric nuclear matter, density dependent interactions
provide a much smaller coexistence region. In the QHD model the instability
region extends to $P\approx0.56$ MeV fm$^{-3}$. The DDRH binodal stretches to
much smaller values of the asymmetry parameter $\xi$ than the one of the
phenomelogical DD-ME2 model. The point of maximum asymmetry is found at
$\xi_{\textmd{MA}}\approx0.05$ and $\xi_\textmd{MA}\approx0.1$ in the DDRH and
DD-ME2 case, respectively. 
We also display the result of the non-relativistic Skyrme type interaction
SLy230a by Dutra et al. \cite{Dutra:2008}. The DD-ME2 binodal is remarkably
close to the one of the SLy230a model. This circumstance can be related to the
way the isovector channel is parametrized in the corresponding models, as was
also found by Dutra et al. To illustrate this more clearly, we show  in
Fig.~\mbox{\ref{fig:Binodal-AsymNM-Models}(b)} the impacts of the
isovector-scalar $\delta$ channel and the momentum correction in the DDRH model
on the binodal shape. The dashed curve represents the result without the
$\delta$ meson interaction, where $\Gamma_\delta$=0.  One can see that the
$\delta$ interaction increases the value of the maximum asymmetry.  The short
dashed line represents the solution without momentum corrections, where we set
all momentum correction parameters ($C_S$) to zero. One can see, that momentum
correction leads to lower values of $\xi_\textmd{MA}$ by approximately the same
extent as of the $\delta$ interaction. It is also interesting to note, that the
binodal section of the DD-ME2 and SLy230a model is very close to the one of the
DDRH model without momentum correction terms.

In conclusion to this discussion, we  study the asymmetry dependence of the
critical temperature.  First of all we should remark that in case of asymmetric
nuclear matter the definition of the critical temperature is used differently in
literature \cite{Ducoin:2006p4509}.  Some authors define the critical
temperature for a given value of $\xi$ such, that $\xi$ corresponds to $\xi_C$
at $T$=$T_C$. In this way the system remains in the gas phase for $T>T_C$
\cite{Zhang:2013p5054}.
On the other hand, many authors prefer the definition which is equivalent to the
one of symmetric nuclear matter, that is, where the pressure has an inflection
point in the $P$-$\rho$ phase diagram.  This definition represents the
temperature, from which the system remains mechanically stable, albeit a
chemical instability may still be present at this point. Therefore, one should
rather refer to this temperature as the critical temperature of mechanical
instability, $T_\textmd{CM}$.  Since the calculation of $T_\textmd{CM}$ is more
straightforward and also more familiar, it is widely used in the literature
\cite{Huber:1998p1295, Sahu:2004p4621, Fiorilla:2012p4093}.  $T_\textmd{CM}$ is
somewhat smaller than $T_C$, yet it also represents the asymmetry dependence of
the instability region.

\begin{figure}[h]
\centering
		\includegraphics{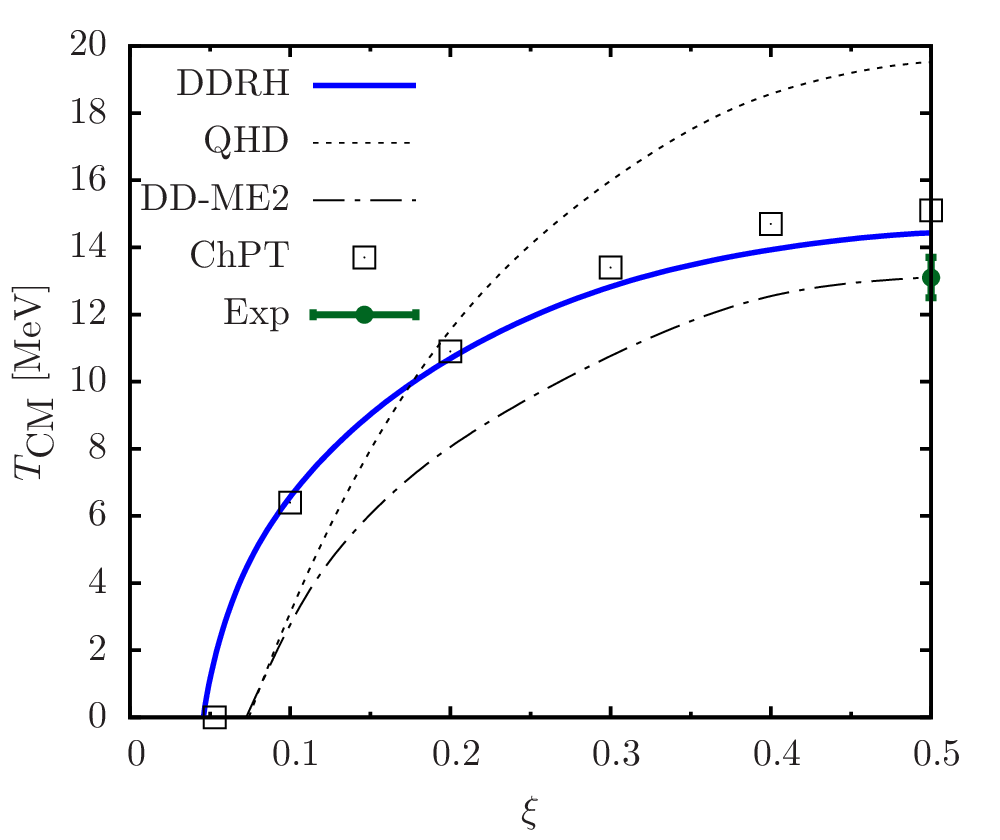}
		\caption{(Color online) The critical temperature of mechanical instability as a function of
			the proton fraction $\xi$. The results are compared between different
			models. The good agreement between our DDRH and the ChPT results of
			\cite{Fiorilla:2012p4093} is remarkable}
		\label{fig:TC-xi}
\end{figure}

In Fig.~\ref{fig:TC-xi} we show the mechanical critical temperature as a
function of $\xi$ for different models.  $T_\textmd{CM}$ decreases continuously
with rising neutron excess and vanishes at $\xi$=$\xi_\textmd{CM}$.  For
asymmetry fractions below this value, the system remains mechanically stable at
all temperatures.  At higher values of $\xi$ the curve of the  QHD model lies
above the curves of the other models, overshooting the experimental value of
symmetric nuclear matter. Nevertheless, the curve falls off much faster with
decreasing $\xi$ and coincides with the DD-ME2 line at $T_\textmd{CM}$=0. 
For comparison also recent results from chiral perturbation theory (ChPT) by
Fiorilla et.al.\  \cite{Fiorilla:2012p4093} are displayed.  In this calculation
the one- and two-pion exchange diagrams as well as $\Delta$-isobar degrees of
freedom are explicitly taken into account. In Table~\ref{tab:TC-xi} we provide a
summary of the results for some fixed values of the proton-neutron asymmetry $\xi$.
It is remarkable that the DDRH result is very close to the one of ChPT 
 throughout the whole $\xi$ range. This indicates that in the DDRH 
the higher order correlation effects on the isospin degree of freedom are implicitly
included in the density dependent terms of the isovector-vector and isovector-scalar channels.

\begin{table}
	\caption{\label{tab:TC-xi} Comparison of the critical temperatures, $T_\mathrm{CM}$ [MeV], at different asymmetry values $\xi$ . }
	\begin{ruledtabular}
		\begin{tabular}{llll}
			$\xi$ & DDRH & DD-ME2 & ChPT \\
			\hline
			0.5 & 14.6 & 13.3 & 15.1 \\
			0.4 & 14.0 & 12.6 & 14.7 \\
			0.3 & 12.8 & 10.8 & 13.4 \\
			0.2 & 10.7 & 8.1  & 10.8 \\
			0.1 & 6.6 & 2.6 & 6.5 \\
	 \end{tabular}
 \end{ruledtabular}
\end{table}

\section{Summary and Outlook}\label{sec:Summary}
The equation of state of asymmetric nuclear matter was studied in the
microscopic DDRH approach. The approach, being based on a Dirac-Brueckner
approach to in-medium interactions, incorporates the essential aspects of an
\emph{ab initio} description by relying only on a free-space
\textit{NN}-interaction but deriving the medium-dependent modifications in a
self-consistent manner. An important step is the projection of the medium
dependence onto effective density dependent meson-baryon interactions leading to
the formulation of relativistic nuclear field theory with  vertex functionals
depending on the field operators. As discussed above, in  mean-field
approximation the field theoretical functionals become functions of the nuclear
density. The mean-field limit is treated as the leading order term in an
expansion around the ground state expectation value of a given configuration of
symmetric or asymmetric nuclear matter. While in former DDRH-related work
investigations of cold nuclear and hypernuclear matter and applications to
neutron stars and finite nuclei and hypernuclei were considered, in this paper
we have applied the approach for the first time to nuclear matter at finite
temperature $T>0$. The nuclear equation of state was investigated in
relativistic mean-field approximation for proton ratios $\xi=Z/A$ ranging from
symmetric nuclear matter ($\xi=\frac{1}{2}$) to pure neutron matter ($\xi=0$). 

As a generic feature of the DDRH theory we have included for the first time also
isovector-scalar fields, realized in nature by the $a_0(980)$ meson. Already by
general symmetry arguments this interaction channel must be included into the
theory. The most important effect is that in asymmetric nuclear matter protons
and neutrons obtain different relativistic effective masses. Hence, the two
species of nucleons become mechanically distinguishable, affecting directly the
thermodynamical properties. Of particular importance is the quite different
behavior of asymmetric matter at the phase boundaries.

Although there exist quite a number of studies on the thermodynamics of
symmetric nuclear matter and pure neutron matter, much less attention has been
paid on the properties of warm asymmetric nuclear matter. While symmetric
nuclear matter can still be considered as a single-component Fermi gas this is
no longer possible in asymmetric nuclear matter. Unequal proton and neutron
numbers are translating into differences in Fermi momenta and chemical
potentials, thus changing the chemical composition and kinetic and, due to
isovector interactions, the mechanical properties of the protons and neutrons,
hence making the two species thermodynamical distinguishable. Thus, as discussed
in the previous sections a much more involved theoretical treatment in terms of
a two-component Fermi gas is required. The increased theoretical effort,
however, is awarded by a much richer phase structure of asymmetric nuclear
matter. As discussed in detail in section \ref{sec:Thermo} the phase structure
of asymmetric nuclear matter is conveniently studied in terms of the total,
isoscalar, and the isovector baryon chemical potentials, respectively,
accounting for the conserved charges of the system, namely the total baryon
number $B$ and the third component $I_3\sim \frac{\rho_n-\rho_p}{\rho}$ of the
total isospin, which is synonym to the conservation of the total charge of the
system. Against na\"{i}ve first expectations, it is not the number of protons
and neutrons separately which is conserved, but the overall Noether-charges are
the conserved quantities. This is seen clearly at the phase transition
boundaries: during the phase transition the baryonic composition might change as
long as $B$ and $I_3$ are conserved. As pointed out, the density dependence and
isospin structure of the in-medium interactions plays a crucial role for the
thermodynamics of warm asymmetric nuclear matter, affecting directly the details
of the phase structure of the system. These aspects were studied in due detail
by comparing our fully microscopic DDRH results with results obtained with the
purely phenomenological RMF-approaches like the original scalar-vector model of
Serot and Walecka and the density dependent extensions as the DD-ME2-approach of
Ring \textit{et al.}. Qualitatively, the three approaches lead to similar
predictions on the thermodynamics of warm nuclear matter, but in detail the
differences in nuclear dynamics are reflected by variations in the phase
diagrams. The positive message of that comparison is that models describing cold
nuclear matter properly are also close in their predictions for warm nuclear
matter, at least in the temperature range below $t\sim 100$~MeV and a
compression factor of up to two or three times nuclear saturation density. This
result is certainly of interest for heavy ion physics because it confirms  and
gives further confidence to the widely used treatment of heavy ion collisions in
terms of a transport theoretical description based on RMF-type dynamics.

Clearly, the approach presented here is open to further extensions. The
inclusion of hyperons is one of the interesting cases allowing to study warm
hypermatter. Adding in addition beta-equilibrium one will be able to describe
warm neutron star matter thus giving access to a more extended, new approach to
the early stages of a neutron star just after the formation of a proto-neutron
star and the subsequent cooling phase.
\vspace{5mm}
\paragraph*{Acknowledgements:}This work was supported in part by DFG
Graduiertenkolleg Giessen-Kopenhagen-Helsinki \textit{Complex Systems of Hadrons
and Nuclei}, GSI Darmstadt, and Helmholtz Graduate School for Hadron and Ion
Research.

\bibliography{Citations}

\end{document}

%% file: definitions.tex
\newcommand{\bfigh}{\begin{figure}[H]}					

\newcommand{\bfig}{\begin{figure}}		

\newcommand{\efig}{\end{figure}}

\newcommand{\hr}{\hat{\rho}}
\newcommand{\al}{\alpha}
\newcommand{\Tr}{\text{Tr}}
\newcommand{\beqn}{\begin{eqnarray*}}
\newcommand{\eeqn}{\end{eqnarray*}}
\newcommand{\beq}{\begin{eqnarray}}
\newcommand{\eeq}{\end{eqnarray}}
\newcommand{\be}{\begin{equation}}
\newcommand{\ee}{\end{equation}}

\newcommand{\vc}[1]{\text{\boldmath $#1$}}

\newcommand{\G}{\Gamma}
\newcommand{\om}{\omega}

\newcommand{\p}{\Psi}
\newcommand{\pb}{\overline{\p}}

\newcommand{\Gh}{\hat\G}

\newcommand{\bmatr}{\begin{pmatrix}}
\newcommand{\ematr}{\end{pmatrix}}
\definecolor{Red}{rgb}{1,0,0}

\newcommand{\td}{\text{d}}

\newcommand{\diff}[2]{{\frac{\partial #1}{\partial #2}}}

\newcommand{\Ld}{\mathcal{L}}

\newcommand{\oover}[1]{\frac{1}{#1}}

\newsavebox{\ZitName}
